\documentclass[10pt,prd,a4paper,superscriptaddress,nofootinbib]{revtex4}
\pdfoutput=1
\usepackage[latin1]{inputenc}
\usepackage[english]{babel}
\usepackage{amsmath}
\usepackage{amssymb}
\usepackage{amsthm}
\usepackage{soul}
\usepackage[normalem]{ulem}
\usepackage[]{graphicx}
\usepackage[]{subfigure}
\usepackage{tensor}
\usepackage{color}
\usepackage{cancel}
\usepackage{fancyhdr}
\usepackage{setspace}
\usepackage[paperwidth=210mm,paperheight=297mm,centering,hmargin=43.7535pt,vmargin=2.5cm]{geometry}
\usepackage[font=small,labelfont=bf,justification=justified]{caption}
\usepackage[bookmarks,linktocpage, colorlinks=true, plainpages = false, citecolor = blue,  linkcolor=blue, urlcolor = blue,
filecolor = blue]{hyperref}
\newcommand{\smallWidthLeft}{262.5pt}
\newcommand{\smallWidthRight}{247.5pt}

\newcommand{\thirdWidthLeft}{175pt}
\newcommand{\thirdWidthRight}{160pt}
\newcommand{\thirdWidthLeftDelayed}{181.667pt}
\newcommand{\thirdWidthLeftDelayedInner}{166.667pt}
\newcommand{\thirdWidthRightDelayed}{171.667pt}
\newcommand{\thirdWidthRightDelayedInner}{156.667pt}
\newcommand{\fullWidth}{390pt}
\newcommand{\hugeWidth}{510pt}
\newcommand{\Vtop}{V_{top}}

\newcommand{\Vplus}{V_{+}}
\newcommand{\Vminus}{V_{-}}
\newcommand{\phiplus}{\varphi_{+}}
\newcommand{\phiminus}{\varphi_{-}}
\newcommand{\phitop}{\varphi_{top}}
\newcommand{\phizero}{\varphi_{0}}
\newcommand{\phibzero}{\bar{\varphi}_0}
\newcommand{\mut}{\mu}

\newcommand{\nt}{N}
\newcommand{\xit}{\xi}
\newcommand{\Htop}{H_{top}}
\newcommand{\Hplus}{H_{+}}

\newcommand{\Vstop}{V''_{top}}
\newcommand{\ntop}{n_{top}}

\begin{document}

\title{Zoology of instanton solutions in flat potential barriers
}

\author{Lorenzo Battarra}
\email{lorenzo.battarra@aei.mpg.de }
\affiliation{Max-Planck-Institut f\"ur Gravitationsphysik, Albert-Einstein-Institut \\
Am M\"uhlenberg 1, D-14476 Potsdam, Germany}

\author{George Lavrelashvili }
\email{lavrela@itp.unibe.ch }
\affiliation{Department of Theoretical Physics,
A.Razmadze Mathematical Institute \\
I.Javakhishvili Tbilisi State University,
GE-0177 Tbilisi, Georgia}

\author{Jean-Luc Lehners $^1$}
\email{jean-luc.lehners@aei.mpg.de}

\date{\today}

\begin{abstract}
We perform a detailed study of the existence and the properties of $O(4)$--invariant instanton solutions in Einstein-scalar theory in the presence of flat potential barriers, i.e. barriers where the second derivative of the potential is small at the top of the barrier. We find a whole zoo of solutions: Hawking-Moss, Coleman--De Luccia (CdL), oscillating instantons, asymmetric CdL as well as other non-standard CdL-like solutions with additional negative modes in their spectrum of fluctuations. Our work shows how these different branches of solutions are connected to each other via ``critical'' instantons possessing an extra zero mode fluctuation. Overall, the space of finite action euclidean solutions to these theories with flat barriers is surprisingly rich and intricate. We find that critical instantons provide the key to understanding both the existence and the properties of instanton solutions.
\end{abstract}
\maketitle

\section{Introduction \label{sec:Intro}}
The problem of tunneling transitions in Einstein--matter theories was first considered in 1980 by Coleman and De Luccia (CdL) in their pioneering article \cite{Coleman1980}. Their analysis revealed that, as in flat space--time \cite{Coleman1977}, the action of so--called \textit{instanton} solutions - $O(4)$--symmetric euclidean solutions respecting appropriate boundary conditions - determines the tunneling rate and, via analytic continuation, the lorentzian evolution of the bubbles of true vacuum produced by tunneling events.
After the CdL instanton another milestone in the investigation of tunneling transitions with gravity
was the discovery of the Hawking-Moss (HM) solution \cite{Hawking:1981fz} which, despite its formal simplicity, raises
many questions of interpretation \cite{Weinberg:2006pc}.

In recent years the study of tunneling in the presence of gravity has gained new momentum. This renewed attention
was triggered on the one side by the string theoretic prediction of a huge landscape of vacua \cite{Bousso2000,Douglas2003}
and, on the other side, by the discovery of new types of ``oscillating'' bounce solutions \cite{Hackworth2005}. 
A thermal derivation of the CdL tunneling prescription was given in \cite{Brown2007}, while in \cite{Brown2011} different situations in which instanton solutions can disappear under small changes in the potential were considered. A large body of work exists by now discussing cosmological applications of instanton solutions, in particular in the context of false-vacuum eternal inflation -- see e.g. \cite{Garriga:1997ef,Garriga:2005av,Freivogel:2005vv,Johnson:2011aa,Lehners:2012wz}.

In the present paper, we study instanton solutions in general relativity minimally coupled to a scalar field with a very flat potential barrier. By this we mean that we consider potentials that have a vanishing or small second derivative at their maximum value; more explicitly we principally consider potentials of the form
\begin{equation}
V(\varphi) = \Vtop - \frac{\mu^2}{2}\varphi^2 - \frac{1}{p}\varphi^p, 
\end{equation}
with  $\mu^2$ small and the two cases $p=4, \, p=6.$  Compact instantons connecting vacua separated by such relatively flat potential
barriers present special properties which were already partially investigated in \cite{Jensen1989,Hackworth2005,Kanno2012,Kanno2012a}. We continue this line of research and give a detailed and systematic investigation of these solutions. 

The flatness of the potential allows for several non--standard solutions, which coexist with CdL and HM instantons. These non--standard instantons have peculiar properties such as additional negative modes and asymmetric profiles despite the symmetry of the potential. Moreover, the dependence of the spectrum of solutions on the parameters of the potential is generally very complicated, and the usual undershooting--overshooting method to find solutions is essentially a blind search. This complexity is the initial motivation for our adoption of \textit{instanton diagrams} as a technique to explore the space of solutions numerically. Instanton diagrams, first introduced in \cite{Bousso2006}, show how each solution evolves with a chosen parameter in the scalar potential, thus describing \textit{branches} of solutions. They allow on the one hand to understand the global structure of the solution space, which can be very complex (for a preview take a look at Figure \ref{fig:phiSix}), and on the other to extract valuable information about the properties of the various instantons. The approach based on instanton diagrams is completed by the computation of both the euclidean action and the number of negative fluctuation modes of the instanton solutions, in order to determine which solution dominates the tunnelling rate, and which solutions contribute to tunnelling at all. 

Our analysis highlights the key role of \textit{critical instantons} in determining both the structure of the space of solutions and the properties of the latter. Critical instantons are solutions which possess an  $O(4)$--invariant zero mode fluctuation, which can be thought of as the cross-over moment of the negative fluctuation mode of one branch of solutions evolving into a positive mode of a different branch. The presence of new branches of solutions can therefore be understood as the consequence of the existence of one or more critical instantons somewhere in the parameter space. Moreover, branches that are separated by a critical instanton have numbers of $O(4)$--symmetric negative modes that differ by one. The simplest example of a critical instanton is the critical HM instanton possessing one negative mode, which separates the CdL branch from the branch of HM instantons with two negative modes. The number of negative modes is crucial in determining the role of instantons as decay--mediating solutions. The fact that, as we show, critical instantons can be geometrically identified on instanton diagrams, makes the latter a very powerful tool in understanding the physical properties of the space of solutions.

Our study shows that the existence of one or more critical instantons with a non--trivial scalar field profile is a generic property of potentials with flat or almost--flat barriers. In fact, we show that some of the known results about oscillating instantons (where the scalar field interpolates more than once between the two sides of the potential barrier), like the equality of the number of oscillations and the number of negative modes \cite{Battarra2012}, are generally valid only in absence of critical instantons other than HM. For example, we uncover solutions that look like ordinary CdL solutions in that they are compact and interpolate only once across the potential barrier, but admit more than one negative fluctuation mode.

The rest of the paper is organized as follows:
In Section \ref{sect:overshooting} we review the basic properties of regular and singular $O(4)$--invariant solutions
in Einstein--scalar field theories, as well as the basic undershooting--overshooting arguments.
In Section \ref{sect:standardCdL} we present our numerical method and apply it to the case of compact CdL instantons.
In Section \ref{sect:fubini} we present numerical results for a flat quartic potential $V = \Vtop- \varphi ^4$.
In Sections \ref{sect:massTerm} and \ref{sect:realistic} we generalize our results to the case of nearly flat,
asymmetric and positive potentials. Finally, in Section \ref{sect:chaotic} we describe the space of instanton
solutions in the highly flat potential $ V = \Vtop - \varphi ^6$. Section \ref{sect:concl} contains concluding remarks.

\section{Instantons in Einstein--scalar field theories \label{sect:overshooting}}
\subsection{Compact and non--compact solutions}

Let us consider the theory of a self--interacting scalar field minimally coupled to gravity
defined by the following Euclidean action
\begin{eqnarray}\label{eq:initialAction}
S_E=\int {d^4x\sqrt{g} \; \Bigl( -\frac{1}{2\kappa}R
+\frac{1}{2}\nabla_\mu\varphi \nabla^\mu\varphi
+ V(\varphi) \Bigr)} \;,
\end{eqnarray}
where $\kappa=8\pi G_{N}$ is the reduced Newton's gravitational constant and the potential $V(\varphi)$ will be specified below.
The general $O(4)$--invariant euclidean Ansatz
\begin{eqnarray}
\varphi & = & \varphi( \eta) \;, \\
ds ^2 & = & d \eta ^2 + \rho ^2( \eta) d \Omega_{3} ^2 \;,
\end{eqnarray}
where $d \Omega_{3} ^2$ is the metric on the unit 3--sphere, is a solution of the theory \eqref{eq:initialAction} when the field equations are satisfied:
\begin{eqnarray} \label{eq:fieldEq1}
\varphi ^{ \prime \prime} & = & - 3 \frac{ \rho'}{ \rho} \varphi ^{ \prime} + V_{, \varphi} \;,\\ \label{eq:fieldEq2}
\rho ^{ \prime 2} & = & 1 +\frac{ \kappa \rho ^2}{3} \left( \frac{1}{2} \varphi ^{ \prime 2} - V \right) \;.
\end{eqnarray}
Note that Eq.~({\ref{eq:fieldEq1}}) has a simple mechanical analogy: it describes the motion of a "particle"
$\varphi(\eta)$ in an inverted potential $-V(\varphi)$ under a frictional force $3 \frac{ \rho'}{ \rho}{ \varphi'}$.

It is easy to show that any solution of (\ref{eq:fieldEq1}, \ref{eq:fieldEq2}) extends to at least one point
where $ \rho = 0:$ defining $\nt\equiv \log{ \rho}$ the field equations read
\begin{eqnarray} \label{eq:scalarFieldx}
\varphi ^{ \prime \prime} & = & - 3 \nt' \varphi' + V_{, \varphi} \;,\\ \label{eq:conservationx}
\nt ^{ \prime 2} - e^{-2 \nt}& = & \frac{ \kappa }{3} \left( \frac{1}{2} \varphi ^{ \prime 2} - V \right) \;.
\end{eqnarray}
Taking a derivative of \eqref{eq:conservationx} and substituting the scalar field equation one gets
\begin{equation} \label{eq:motionx}
\nt ^{ \prime \prime} = - e^{-2 \nt} - \frac{ \kappa \varphi ^{ \prime 2}}{2} \;.
\end{equation}
If we consider \eqref{eq:motionx} as a one--dimensional equation of motion, the particle located at $\nt( \eta)$ is subject to a potential
$v(\nt) = - e^{-2\nt}/2$ and a time--dependent force pushing it towards negative $\nt$: the ``potential'' is then steep enough for $\nt$ to reach
$N = - \infty$ i.e. $ \rho = 0$ in a finite ``time'', either in the past or the future. The general solution could fail extending to these points if
a singularity showed up at a finite value of $ \rho$ but, under the assumption that the potential has no singularity at finite $ \varphi$,
this can be proven not to occur (see Appendix \ref{app:one}). Redefining the sign and the origin of $ \eta$ one can then assume
$ \rho( \eta = 0) = 0$, and $ \rho > 0$ for $\eta$ positive. Regularity at $ \eta = 0$ implies the boundary conditions
\begin{eqnarray} \label{eq:bcphi}
\varphi'(0) & = & 0 \;,\\
\rho'(0) & = & 1 \;, \label{eq:bcrho}
\end{eqnarray}
Depending on the value of $ \phizero \equiv \varphi( \eta = 0)$ and on the shape of $V( \varphi)$, a solution satisfying these
boundary conditions can be \textit{compact} or \textit{non--compact} \cite{Bousso2006}:
\begin{itemize}
\item \textbf{compact solutions}: $ \rho( \eta) $ reaches a maximum $ \rho_{m}$ and then returns to $ \rho = 0$
at some finite ``time'' $ \bar{  \eta} >0$. The existence of such solutions requires $V$ to be strictly positive \textit{somewhere}
in field space. Indeed, in this case $\rho$ must have a local maximum at some intermediate value of $ \eta$ and  \eqref{eq:fieldEq2}
requires $V>0$ there.
\item \textbf{non--compact solutions}: $ \rho \rightarrow \infty$  monotonically as $ \eta \rightarrow \infty$. The existence of
non--compact solutions requires the potential to be negative or zero somewhere. Indeed, in this case $N' > 0$ everywhere and the
quantity on the rhs of \eqref{eq:conservationx} is monotonically decreasing from its initial value $ -\kappa V( \phizero)/3$
because of the friction term in \eqref{eq:scalarFieldx}. If the solution is non--compact the lhs of \eqref{eq:conservationx}
cannot approach a negative constant, so $ V( \phizero) \leq 0$. In fact, $V( \phizero) = 0$ is compatible with a
non--compact solution only when $ \varphi( \eta) = \phizero$ solves the scalar field equation i.e. when $ \phizero$ is a
stationary point of $V$, in which case the corresponding instanton is four--dimensional flat euclidean space.
\end{itemize}
Among the one--parameter family of solutions of the field equations obeying \eqref{eq:bcphi}, \eqref{eq:bcrho},
the ones which may be relevant for tunneling are those which respect the same boundary conditions as the false
vacuum euclidean geometry, whose action enters in the expression of the decay rate
\begin{equation}
\Gamma \propto \textrm{exp}\left\{-  \left(S_E( \varphi) - S_E( \varphi_{fv}) \right) \right\} \;.
\end{equation}
These special solutions of the field equations are referred to as \textit{instantons}. Compact solutions, which describe
tunneling from de Sitter space and will be the focus of this paper, have the topology of a four--sphere whose
``north pole'' can be taken to be $ \eta = 0$. In this case, the natural boundary condition is the requirement of regularity
at the ``south pole'' $ \rho( \bar{ \eta}) = 0$ of the euclidean geometry:
\begin{eqnarray} \label{eq:bcphi2}
\rho'( \bar{ \eta}) & = & - 1 \;,\\ \label{eq:bcrho2}
\varphi'( \bar{ \eta}) & = & 0 \;.
\end{eqnarray}
When the scalar potential is everywhere well--defined, these conditions can be shown to be equivalent to requiring a finite
limit $ \phibzero \equiv \lim_{ \eta \rightarrow \bar{ \eta}} \varphi( \eta)$ for the scalar field (see Appendix \ref{app:two}).
Unlike for the non--compact case, for compact instantons the scalar field cannot approach a stationary point of $V$
at the south pole $ \eta = \bar{ \eta}$. Thus $ \phibzero \equiv \varphi( \bar{ \eta})$ is always different
from the false vacuum value and, as described in \cite{Brown2007}, a thermal fluctuation from the minimum of the potential to $\phibzero$ is needed to initiate tunneling from de Sitter space.
Moreover, the similarity of the boundary conditions (\ref{eq:bcphi2}, \ref{eq:bcrho2}) and
(\ref{eq:bcphi}, \ref{eq:bcrho}) and the invariance of the field equations under $ \eta \rightarrow \bar{ \eta} -\eta$ imply that if
$ \phizero$ corresponds to a regular solution so does $ \phibzero$.
\subsection{Overshooting and undershooting}
The existence of instanton solutions is usually proven through the so--called \textit{overshooting--undershooting} argument
\cite{Coleman1980}. For each value of $ \phizero \equiv \varphi( \eta = 0)$, a unique solution of the field equations can be
found with initial conditions \eqref{eq:bcphi}, \eqref{eq:bcrho}. For generic values of $ \phizero$, however, the solution will not
respect the boundary conditions at $ \eta = \bar{ \eta}$. In the compact case, this means that generically
$ \varphi \rightarrow \pm \infty$ as $ \eta \rightarrow \bar{ \eta}$. Across a discrete set of values
$ \{\phizero ^{i}\}$ however, the sign of this divergence changes, e.g.:
\begin{eqnarray} \label{eq:behaviorOne}
&\varphi \stackrel{ \eta \rightarrow \bar{ \eta}}{ \longrightarrow} + \infty, &\quad \phizero < \phizero ^{i} \;,\\ \label{eq:behaviorTwo}
&\varphi \stackrel{ \eta \rightarrow \bar{ \eta}}{ \longrightarrow} - \infty, &\quad \phizero > \phizero ^{i} \;.
\end{eqnarray}
From the continuous dependence of the solution on the parameter $ \phizero$
one deduces that the solution corresponding to each separating value $ \phizero ^{i}$ cannot have either divergent
behavior, hence it is necessarily regular. A transition from \eqref{eq:behaviorOne} to \eqref{eq:behaviorTwo} is always
associated with a change in the \textit{number of oscillations} or \textit{number of passes} $n$ of the scalar field,  conventionally defined as the number
of zeros of $ \varphi'$ plus one (excepting the boundary condition zero $ \varphi'(0) = 0$ and the possible one at
$ \eta = \bar{ \eta}$), in such a way that an instanton
with a monotonically varying scalar field has $ n = 1$. In the compact case,
instantons appear at those special values of $ \phizero$ across
which the discrete function  $ n( \phizero)$ has a jump. In particular, based on the analysis of perturbations of the regular solutions,
one can show that $ n$ can only jump by a unit value \cite{Bousso2006}: a transition like \eqref{eq:behaviorOne} \eqref{eq:behaviorTwo}
is necessary and sufficient for the existence of an instanton
solution, whose number of oscillations is $ \textrm{min}\{n( \phizero ^{i} - \epsilon), n( \phizero ^{i} + \epsilon)\}$
\footnote{The non--compact case can be very different. For example, $ n( \phizero)$ can jump by more than unity
\cite{Bousso2006}. Moreover, depending on the shape of the potential and on the boundary conditions
at $ \eta = \bar{ \eta} = \infty$, the existence of instanton solutions may not require any jump in $ n( \phizero)$.}.

\begin{figure}[t]
\begin{minipage}{\thirdWidthLeft}
\flushleft
\includegraphics[width=\thirdWidthRight]{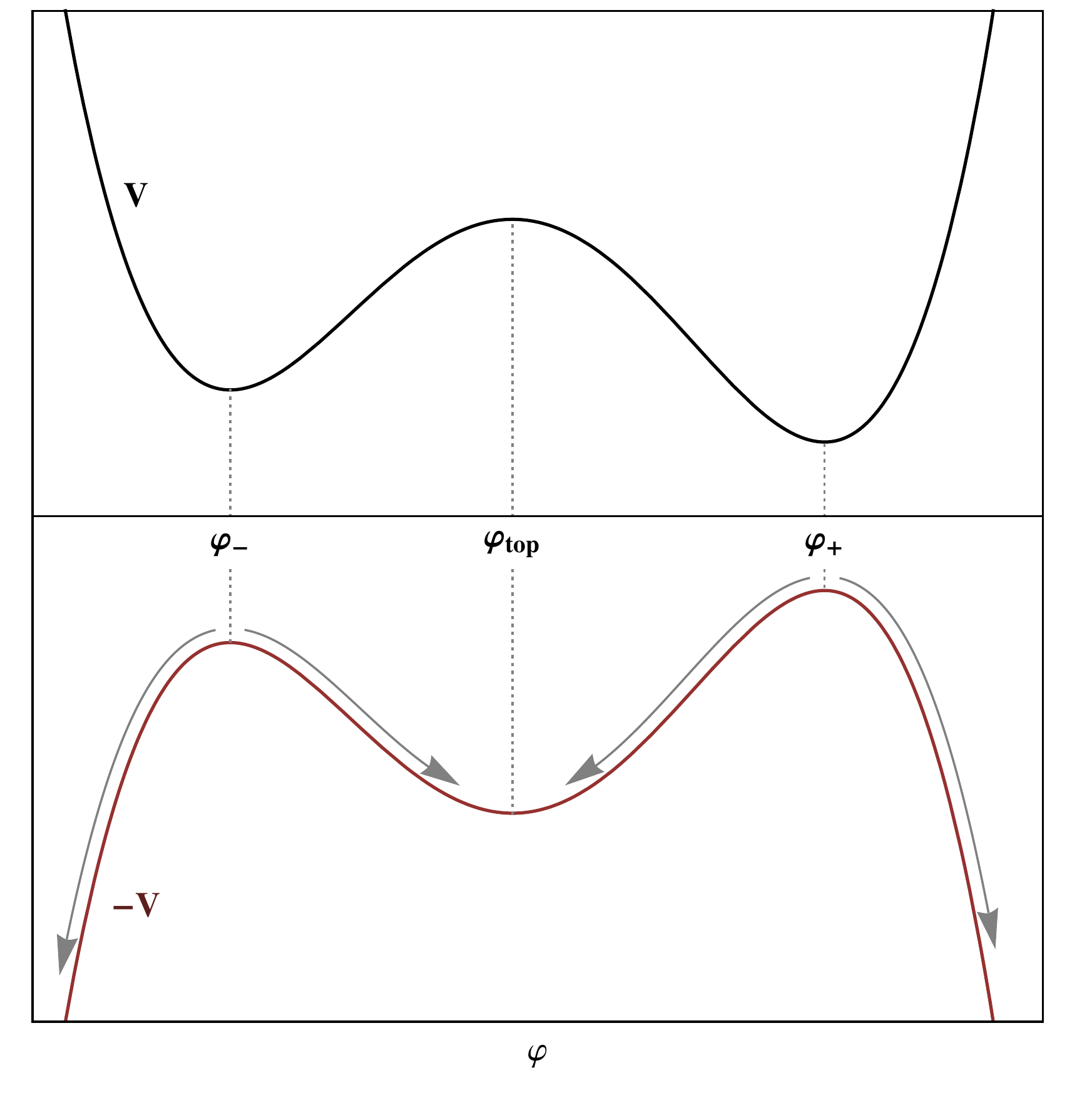}
\end{minipage}%
\begin{minipage}{\thirdWidthRight}
\flushleft
\includegraphics[width=\thirdWidthRight]{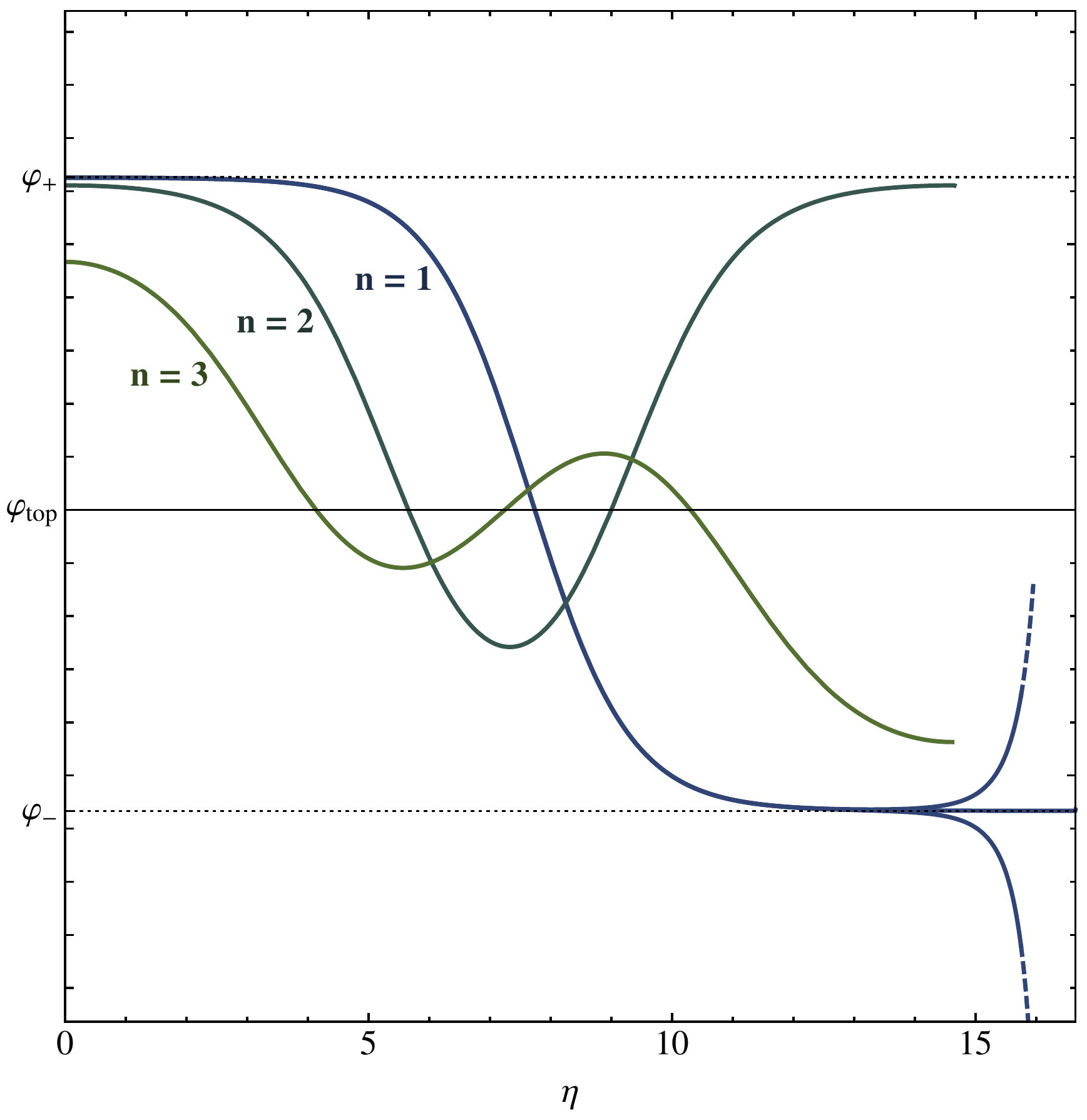}
\end{minipage}
\begin{minipage}{\thirdWidthRight}
\flushleft
\includegraphics[width=\thirdWidthRight]{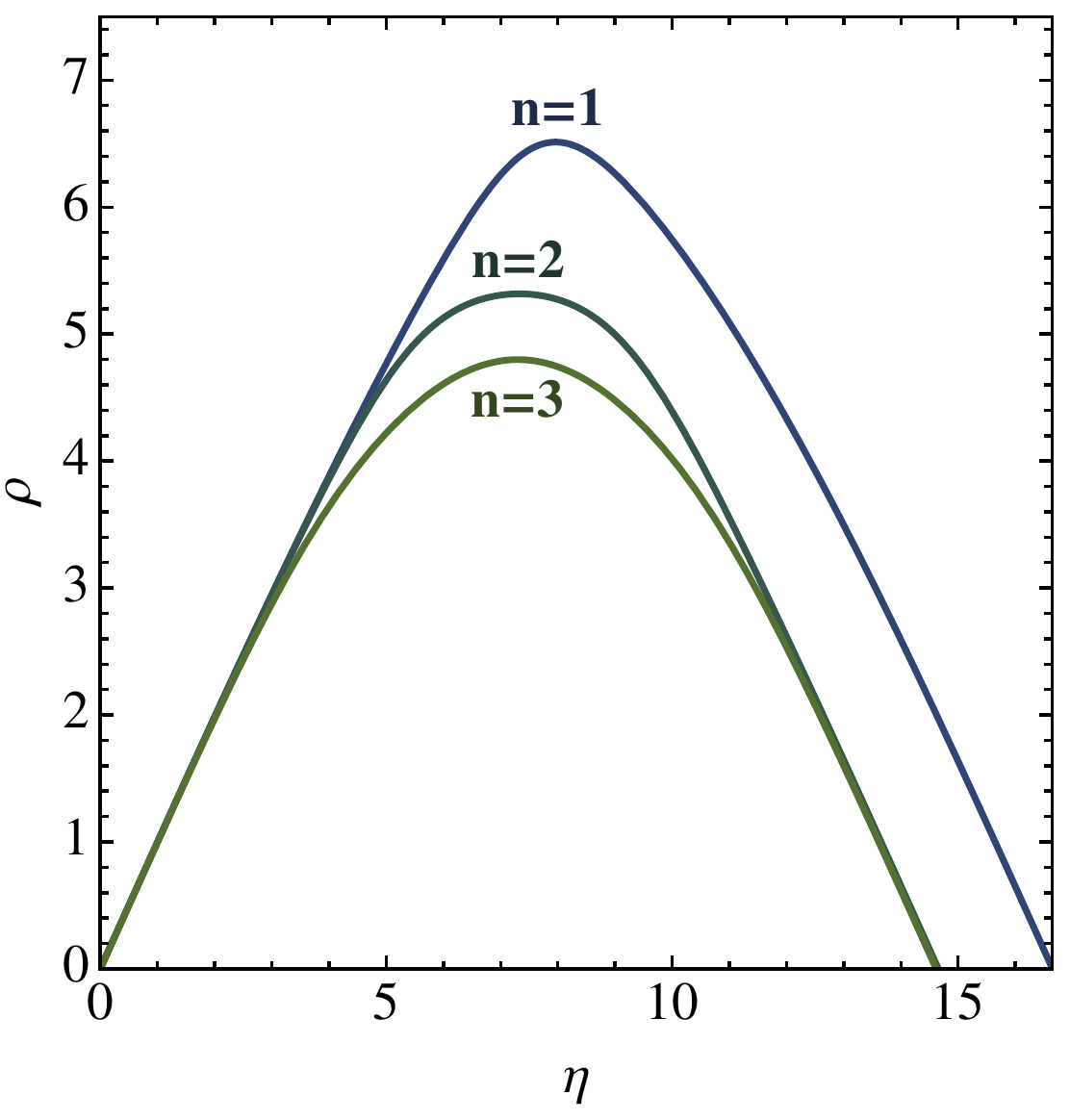}
\end{minipage}
\caption{\small \label{fig:examplePotentialAndExampleOvershooting}Left panel: double well--potential
and its euclidean version (bottom). Center and right panels: $( \varphi, \rho)$ profiles of three instanton solutions for the potential \eqref{eq:potential234},
with $ \xit =  0.1$, $\Vtop = e ^{-2}$ and $\Vtop/\Vplus = 4$.  The appearance of the $n=1$ as separating solution
between $n=1$ and $n=2$ is highlighted (see also Figure \ref{fig:exampleMapAndExampleEuclideanAction}).}
\end{figure}

Overshooting--undershooting arguments prove the existence of jumps in $n( \phizero)$, hence the existence of an
instanton solution, by showing that $n( \phizero)$ takes different values at special points in field space. The simplest
example of such an argument applies to vacuum decay in flat--space \cite{Coleman1977}. Consider a simple double--well
potential as in Figure \ref{fig:examplePotentialAndExampleOvershooting}. The equation for the scalar field reads
\begin{equation}
\varphi'' + \frac{3}{ \eta} \varphi' = V'( \varphi) \;.
\end{equation}
This is the equation for a particle subject to a potential $-V$ and a friction force with coefficient $3/ \eta$. If $ \phizero$
is set close enough to $ \phitop$, the scalar field will not escape from the euclidean potential well and will
undergo an infinite number of damped oscillations, hence $ n(  \phitop + \epsilon) =  \infty$ (\textit{undershooting}).
On the other hand, if $ \phizero$ is set very close to the local maximum $ \phiplus$ of the inverted potential,
the scalar field will escape from it at arbitrarily large $ \eta$:
\begin{equation} \label{eq:escapeTime}
\eta_{escape} \gtrsim - \frac{1}{|V''( \phitop)| ^{1/2}} \log \left( \phiplus - \phizero \right) \;.
\end{equation}
In this way, the friction coefficient can be made arbitrarily small during the non--trivial part of the scalar field
evolution, and provided
\footnote{Our notation should be obvious: $V(\varphi_{\pm}) \equiv V_{\pm}, V(\phitop) \equiv V_{top}$, and $V( \phizero)  \equiv V_{0}$.  }
$\Vplus < \Vminus$ the scalar field can go beyond $ \phiminus$ and then diverge as the potential
increases again (\textit{overshooting}). Therefore, $ n( \phizero =  \phiplus - \epsilon) = 1$ and an $n=1$ instanton
must separate the undershooting and overshooting regimes.
\section{Standard compact CdL solutions \label{sect:standardCdL}}
\subsection{Overshooting and undershooting in de Sitter space}
When gravity is taken into account and the potential allows only compact instantons, as the one in
Figure \ref{fig:examplePotentialAndExampleOvershooting}, both ends of the undershooting--overshooting argument need to be updated.
All solutions being compact, the friction coefficient $ \rho'/ \rho$ becomes negative before $ \eta = \bar{ \eta}$ (\textit{anti--friction}) and the scalar field diverges
except when the solution is regular. Therefore, for $ \phizero$ very close to $ \phitop$ the number of oscillations
$ n( \phizero)$ remains finite. Its value depends on the shape and value of the potential at the top of the barrier
$ \varphi = \phitop$, and was determined \cite{Hackworth2005} to be the smallest integer $ \ntop$ satisfying
\begin{equation} \label{eq:bounddS}
\ntop(\ntop+3) > \frac{|\Vstop|}{\Htop ^2}, \quad \Htop \equiv \sqrt{ \frac{ \kappa \Vtop}{3}} \;.
\end{equation}
This relation translates the fact that only a finite number of oscillations around $ \phitop$ can be fit in the Hubble
radius $\Htop ^{-1}$ of the compact geometry, before anti--friction kicks the scalar field away from the top of the potential barrier.
At the other end of field space, namely for $ \varphi = \phiplus - \epsilon$, the overshooting argument applies in a stronger
form than in flat space: even if $\Vplus\geq \Vminus$, one finds $ n( \phizero \lesssim \phiplus) = 1$ provided $\Vplus > 0$.
Indeed, if the scalar field is initially set very close to $ \phiplus$, the naively expected escape time \eqref{eq:escapeTime} can
exceed the Hubble radius $\Hplus ^{-1}$. In this case the scalar field is kicked away from $ \phiplus$ by the divergent anti--friction
force: by decreasing $ (\phiplus - \phizero)$ this kick can be made large enough to let the scalar field go over $ \phiminus$
and diverge monotonically. This means that $n( \phizero)$ varies between $\ntop$ and $1$ as $ \phizero$ varies
between $ \phitop$ and $ \phiplus$. Hence, an odd number of instantons with $ \phitop < \phizero < \phiplus$ exists for each value
of $n$ between $1$ and $\ntop-1$ (including both extrema). The same conclusion clearly applies to the part of the potential between
$ \phiminus$ and $ \phitop$.

Finding the instanton solutions amounts to finding those special values of $\phizero$ across which the function $n( \varphi_0)$ jumps by unity. In most cases, the function $n( \varphi_0)$ can only be evaluated by numerically solving the field equations with the boundary condition $ \varphi( \eta = 0) = \phizero$ and counting the number of oscillations in the corresponding scalar field profile. Then, if $n( \phizero)$ is known to jump by unity inside a given interval, the standard method consists in finding the approximate position of the discontinuity by consecutive bisections. The main difficulty of this procedure consists in roughly detecting \textit{all} the discontinuities of $n(\phizero)$ when the initial bisection intervals are chosen. This difficulty arises when the position and sign of the discontinuities do not follow a simple pattern or, more precisely, when this pattern exhibits a complicated dependence on the parameters of the potential. As it will be illustrated in the next section, instanton diagrams simplify the task by providing a direct overview of this dependence, which can be then used as a guide to find single solutions.
\begin{figure}[t]
\begin{minipage}{\smallWidthLeft}
\flushleft
\includegraphics[width=\smallWidthRight]{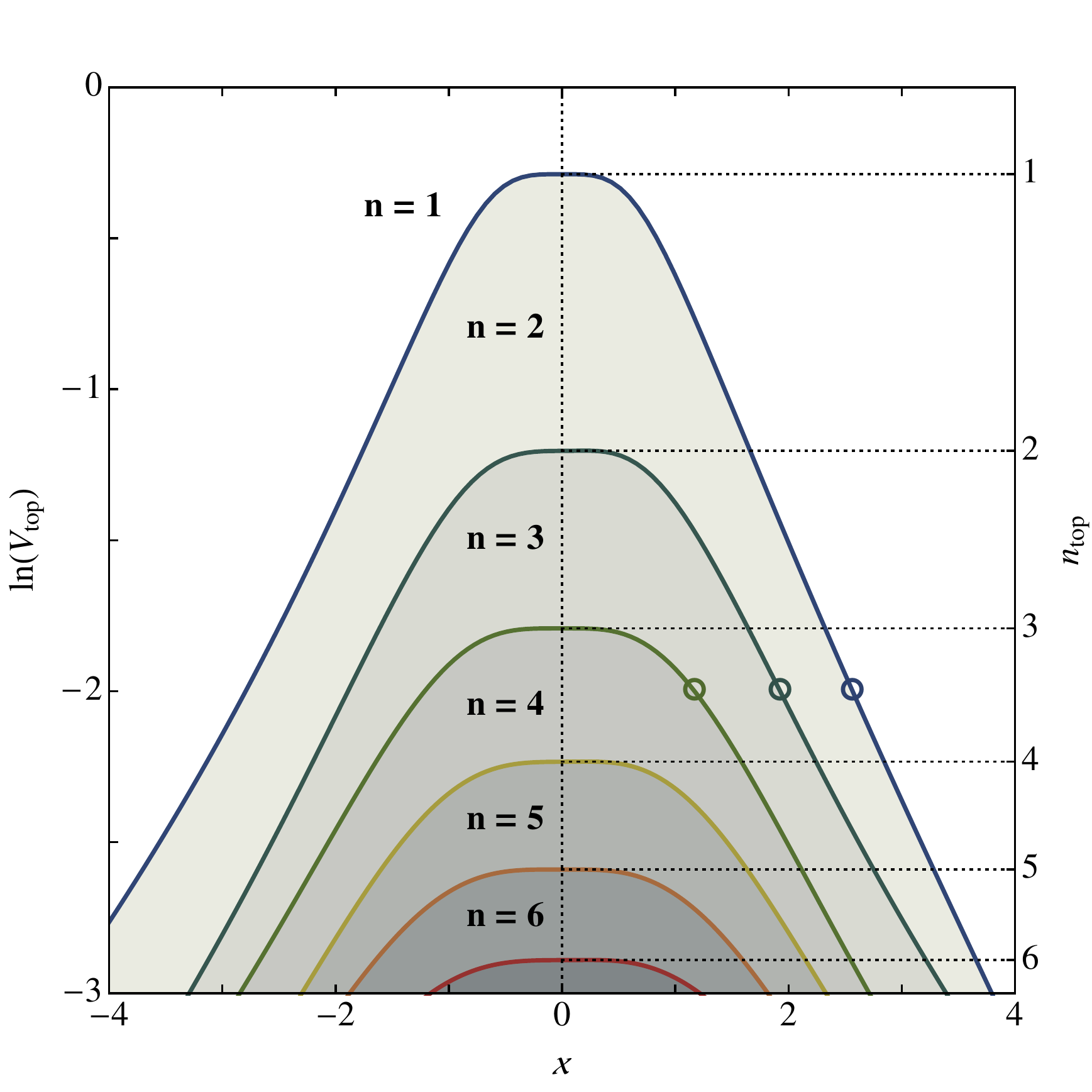}
\end{minipage}%
\begin{minipage}{\smallWidthRight}\flushleft
\includegraphics[width=\smallWidthRight]{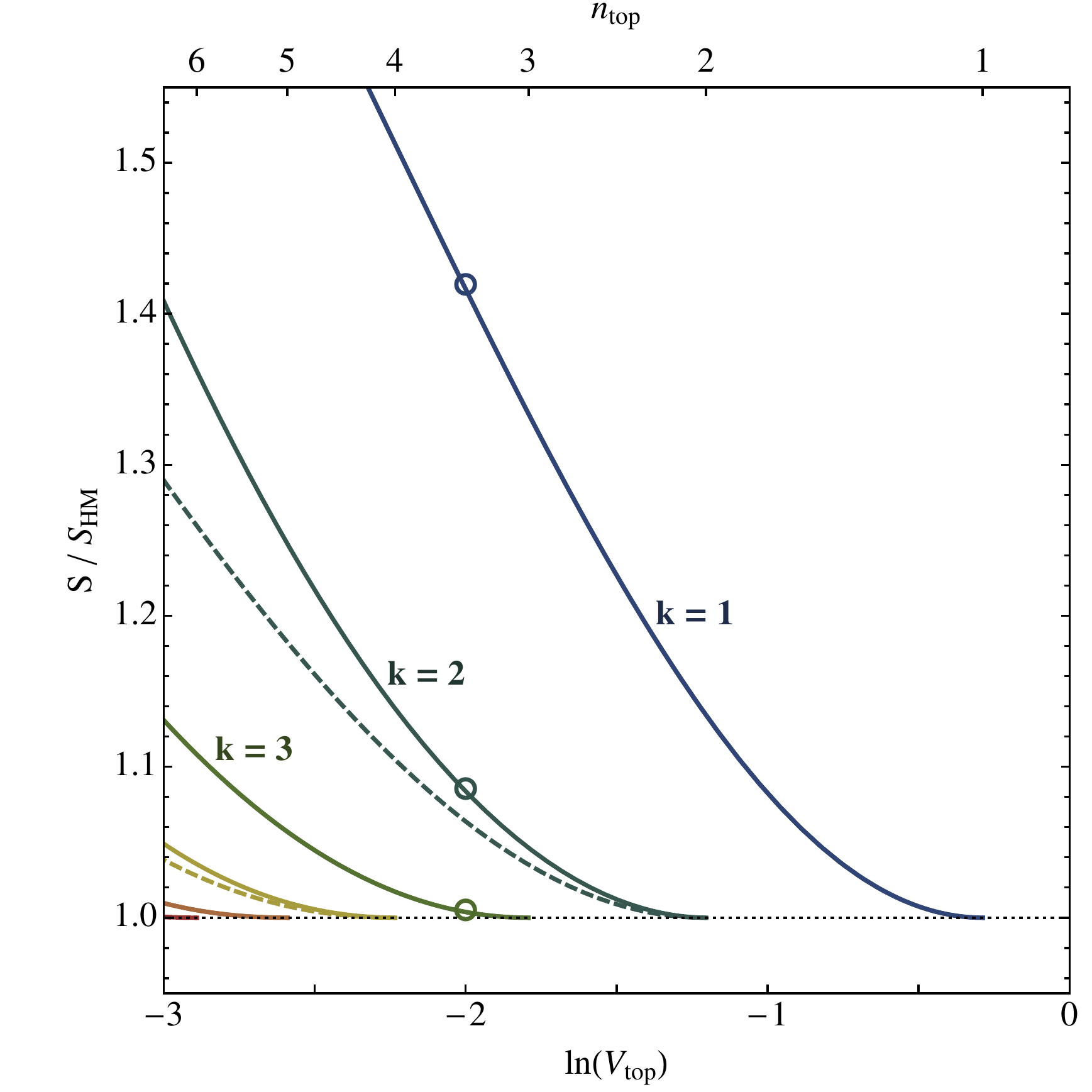}
\end{minipage}\caption{\small \label{fig:exampleMapAndExampleEuclideanAction}Left panel: instanton diagram
for the potential \eqref{eq:potential234} with $ \xit = 0.1$ and $\Vtop/\Vplus = 4$. The three circles correspond to the solutions
represented in the right panel of Figure \ref{fig:examplePotentialAndExampleOvershooting}. The vertical, dotted line represent $\phizero = 0$ and the continuous family of HM solutions: the value of $n$ on this line is formally undetermined but can be interpolated from the neighboring solutions. Right panel: euclidean action of
the different branches of instanton solutions normalized to the action of the corresponding Hawking--Moss instanton.
Solid lines (resp. dashed lines) represent the instantons with $x>0$ (resp. $x<0$). The number of negative modes $k$ is also indicated.}
\end{figure}
\subsection{Instanton diagrams \label{ssect:diagrams}}
In many cases, \textit{exactly} one instanton exists for every $n$ between $1$ and $\ntop-1$. Put differently, $n(\phizero)$ jumps only by \textit{positive} unities when $ \varphi_0$ is varied from $\phiplus$ to $\phitop$. As an example, consider the potential:
\begin{equation} \label{eq:potential234}
V = \Vtop - \frac{1}{2} \varphi ^2 - \frac{ \lambda ^{1/2} \xit}{3}\, \varphi ^3 + \frac{ \lambda}{ 4} \varphi ^4 \;.
\end{equation}
The left panel of Figure \ref{fig:exampleMapAndExampleEuclideanAction} depicts an \textit{instanton diagram},
with the $x$ variable defined by:
\begin{equation} \label{eq:xdef}
\phizero = \left\{\begin{array}{cl}
 \phiplus \left(1 - e^{-x ^2} \right), & x > 0 \;,\\
\phiminus \left( 1- e^{-x ^2} \right), & x < 0 \;.
\end{array}
\right.
\end{equation}
The use of this logarithmic variable is particularly helpful because instanton initial values $\phizero$ tend to
accumulate near the vacuum values $\varphi_{\pm}$. The parameter $ \lambda$ in the potential is adjusted to keep the ratio $r =\Vplus/\Vtop$ constant and equal to $1/4$, and we set  $ \xit = 0.1$, $ \kappa = 1$. 

Each point of the diagram specifies a potential (via $\Vtop$) and a boundary condition (via $\phizero$) which can be used to solve the field equations, and extract the number of oscillations $n(\phizero,\Vtop)$ of the scalar field. The instanton diagram shows the regions of constancy of $n(\phizero,\Vtop)$ and the \textit{instanton curves} across which the function is discontinuous. According to the discussion of the previous section, each point $(\phizero ^{*}, \Vtop ^{*})$ on a curve separating the $n= n ^{*}$ from the $n = n ^{*}+1$ regions corresponds to an instanton with $ n ^{*}$ oscillations in the theory specified by the parameter $\Vtop ^{*}$. In other words, the instanton solutions of a given theory are the intersection of the instanton curves with the horizontal line corresponding to the theory. Globally, the various curves describe how a given instanton evolves when the potential is changed. In order to obtain the instanton diagram numerically, the function $n(x, \Vtop)$ is sampled on a grid of values of $( x, \Vtop)$. A more accurate estimate for the values of  $\phizero$ corresponding to particular instantons can then be obtained by using a bisection algorithm as explained in the previous section. 

The diagram in Figure \ref{fig:exampleMapAndExampleEuclideanAction} shows that, for fixed $\Vtop$, the function $n(\phizero)$ indeed grows monotonically from $n=1$ to $n=\ntop$ when $\phizero$ is varied from $\phiplus$ ($x = \infty$) to $\phitop$ ($x=0$), so that a single instanton is found for each $n=1, \ldots, \ntop - 1$. A similar conclusion holds for $\phiminus < \phizero < \phitop$. It should be rather clear that, in cases like the one depicted in Figure \ref{fig:exampleMapAndExampleEuclideanAction}, the instanton diagram technique only brings a different way to visualize the location of various solutions in field/parameter space. Indeed, the appearance and the evolution of the different branches can also be studied by simply studying a few constant $\Vtop$ slices, as done in \cite{Battarra2012}. Moreover, the fact that $n(\phizero)$ varies monotonically makes the search for its discontinuities a rather straightforward task. On the other hand, in the case of flat potential barriers this visualization is extremely helpful in
understanding the nature of various non--standard solutions, which are the subject of the next sections.

\subsection{Zero and negative modes}
Let's consider again the global structure of the instanton diagram in Figure \ref{fig:exampleMapAndExampleEuclideanAction}. As $\Vtop$ decreases, $\ntop$ (which we defined in \eqref{eq:bounddS}) increases and more and more instanton solutions appear,
moving away from the Hawking--Moss (HM) solutions $ \varphi = \phitop$ located at:
\begin{eqnarray} \label{eq:crhm1}
V_n & = & \frac{|\Vstop|}{ \kappa} \frac{3}{n(n+3)} \;, \\ \label{eq:crhm2}
\rho_n & = & H _{n} ^{-1} \, \sin{(H _{n} \eta)}, \quad H _{n} = \sqrt{ \frac{ \kappa V_n}{3}} \;.
\end{eqnarray}
For $\Vtop$ smaller but infinitesimally close to $V_n$, the $n$--oscillating instantons have infinitesimally small $\phizero$ and appear as a regular, $O(4)$--invariant perturbation (or \textit{zero mode}) of the HM instanton (\ref{eq:crhm1},\ref{eq:crhm2}) which, according to Section \ref{sec:Intro}, is called \textit{critical}. This zero mode is the deformation of the critical solution produced by an infinitesimal variation of the boundary condition $\phizero$, i.e. an infinitesimal horizontal displacement in the instanton diagram. The regularity of this deformation is signalled by the horizontal slope of the instanton curve: the infinitesimal variation of $\phizero$ transforms HM into an infinitesimally close solution that still lies on the curve, i.e. the transformed solution is also a regular solution to the same theory. This is true more generally: \textit{a critical instanton corresponds to a point in the diagram where the tangent instanton curve is parallel to the $\phizero$ axis}.

It follows from this observation that all the instantons located on a single curve in the diagram of Figure
\ref{fig:exampleMapAndExampleEuclideanAction} possess the same number of \textit{negative modes} in the $O(4)$--invariant sector
of the scalar field perturbations. Denoting the gauge-invariant scalar fluctuation mode by $f,$ it satisfies the equation  \cite{Khvedelidze2000,Lavrelashvili2000,Gratton2001,Lavrelashvili2006,Dunne:2006bt} 
\begin{equation} \label{eq:pertEquation}
- f'' + U[ \rho( \eta), \varphi( \eta)] f = \epsilon\,f \;,
\end{equation}
with eigenvalue $\epsilon$ and effective potential
\begin{eqnarray}
U[\rho(\eta), \varphi(\eta)] & \equiv &  \frac{1}{ \mathcal{ Q}} V_{, \varphi \varphi} - \frac{10 \rho ^{ \prime 2}}{ \rho ^2 \mathcal{Q}}
+ \frac{ 12 \rho ^{ \prime 2}}{ \rho ^2 \mathcal{ Q} ^2} + \frac{8}{ \rho ^2 \mathcal{Q}} - \frac{ 6}{ \rho ^2}
- \frac{ 3 \mathcal{Q}}{ \rho ^2} - \frac{ \rho ^{ \prime 2}}{4 \rho ^2} \nonumber \\
&&+ \frac{ \kappa \rho ^2}{2 \mathcal{Q} ^2} V _{, \varphi}^2 - \frac{ 2 \kappa \rho \rho' \varphi'}
{ \mathcal{Q} ^2} V_{, \varphi} - \frac{ \kappa}{6} \left( \varphi ^{ \prime 2} + V \right) \;,\\
\mathcal{Q} & \equiv & 1 - \frac{ \kappa \rho ^2 \varphi ^{ \prime 2}}{6} \;.
\end{eqnarray}
A property of this equation is that the eigenvalues $\{ \epsilon ^{i}\} _{i=1,2, \ldots}$ of different families of solutions vary continuously on each family. Therefore, additional negative eigenvalues can only appear at critical points
corresponding to instantons possessing a \textit{zero mode}, i.e. a regular perturbation mode. For the same reason, the number of negative modes along an instanton curve generally increases or decreases by one across a critical solution. In the case described in Figure \ref{fig:exampleMapAndExampleEuclideanAction}, the number $k$ of $O(4)$--invariant
negative modes of oscillating instantons ($\phizero > 0$) turns out to be equal to the number of oscillations (see \cite{Battarra2012} for
a detailed study of these negative modes), i.e. $k = n$. On the other hand, the number of negative modes of HM instantons is always equal to $\ntop$, and jumps from $n$ to $n+1$ across the critical solution (\ref{eq:crhm1},\ref{eq:crhm2}). Hence, in agreement with the discussion of Section \ref{sec:Intro}, the critical HM solution at $\Vtop = V_n$ separates a branch of solutions with $n$ negative modes, namely the $n$--oscillating instantons, from a branch with $n+1$ negative modes, namely the HM solutions with $V_{n+1} < V < V_{n}$. In a sense, the criticality of the HM solutions explains the existence of the standard oscillating instantons.

Finally, as shown in the right panel of Figure \ref{fig:exampleMapAndExampleEuclideanAction},  the euclidean action of the $n=1$ solutions turns out to be always
 the most negative and, in particular, to be more negative than the action of the corresponding HM solutions,
 \begin{equation}
 S_{HM} = -\frac{24 \pi ^2}{ \kappa ^2 \Vtop} \;.
 \end{equation}
Therefore the $n=1$ instantons are the solutions which determine the tunneling rate in the theory specified by the
potential \eqref{eq:potential234}, and they are commonly referred to as Coleman--de Luccia instantons.


\section{Tunneling through flat barriers: quartic potential \label{sect:fubini}}
The case we just described is the simplest one compatible with the values of $n(\phizero)$ obtained from the
undershooting--overshooting arguments: $n(\phizero)$ grows monotonically from $1 = n(\phiplus - \epsilon )$ to $\ntop = n(\phitop  + \epsilon)$.
However, sufficient conditions on the shape of the potential are not known which guarantee this behavior of $ n( \varphi_0)$. In
particular, when $\ntop = 1$ the number of instantons of any order $n$ with $\phitop < \phizero < \phiplus$ is constrained to
be an \textit{even} integer, but the undershooting--overshooting argument does not force it to vanish. The corresponding
potentials are characterized by a \textit{flat} barrier separating the two vacua:
\begin{equation}
\frac{3|\Vstop|}{ \kappa \Vtop} \leq 4 \;.
\end{equation}
A particular case of exactly flat potential was studied in \cite{Jensen1989}, where the existence of several $n=1$ solutions
was established despite $\ntop = 1$. In particular, it was shown that these solutions have more negative euclidean action than
the corresponding HM solution. In \cite{Hackworth2005} similar results were found together with other non--standard
behaviors of the $n( \phizero)$ function. Moreover, the authors suggested the existence of
a generalized bound on the curvature of the potential \textit{around} the top of the barrier for the existence of $n=1$ instantons.
However, the only analytic result known so far is a \textit{necessary} condition, stating that the inequality
\begin{equation}
\frac{3|V''( \varphi)|}{ \kappa V( \varphi)} > 4 \;,
\end{equation}
must be satisfied for some value of $ \varphi$ in order for $n=1$ solutions to exist \cite{Balek2004}.

In order to reconsider the question of tunneling through flat barriers, we start from the simple quartic potential
\begin{equation} \label{eq:potentialFubini}
V = \Vtop - \frac{ \lambda}{4} \varphi ^4, \quad \lambda > 0 \;.
\end{equation}
In flat space ($ \kappa = 0$), the conformal invariance of the scalar field theory allows for the existence of a continuous
family of instanton solutions \cite{Fubini:1976jm}, called \textit{Fubini} instantons, for which the scalar field
approaches $ \phitop = 0$ asymptotically:
\begin{equation}
\varphi_b( \eta) = \sqrt{ \frac{8}{ \lambda}} \frac{ b}{ \eta ^2 + b ^2}, \quad b \in \mathbb{R} \;.
\end{equation}
In fact S.~Fubini found these finite action euclidean solutions in scalar
field theory with conformally invariant (quartic) potential
almost at the same time as instantons were discovered in the Yang-Mills theory \cite{Belavin:1975fg}.
Later Fubini instantons were rediscovered several times \cite{Affleck:1980mp,Lee:1985uv}.
They describe tunneling without barrier \cite{Lee:1985uv,Lee:1986saa,Linde:2005ht},
and have applications in particle physics as well as in the context of the AdS/CFT conjecture
\cite{Barbon:2010gn,Barbon:2011ta,deHaro:2006nv,deHaro:2006wy}.
Moreover, it was suggested \cite{Loran:2006sf} that Fubini vacua can be used as classical de Sitter vacua.

\begin{figure}[t]
\begin{minipage}{\smallWidthLeft}
\flushleft
\includegraphics[width=\smallWidthRight]{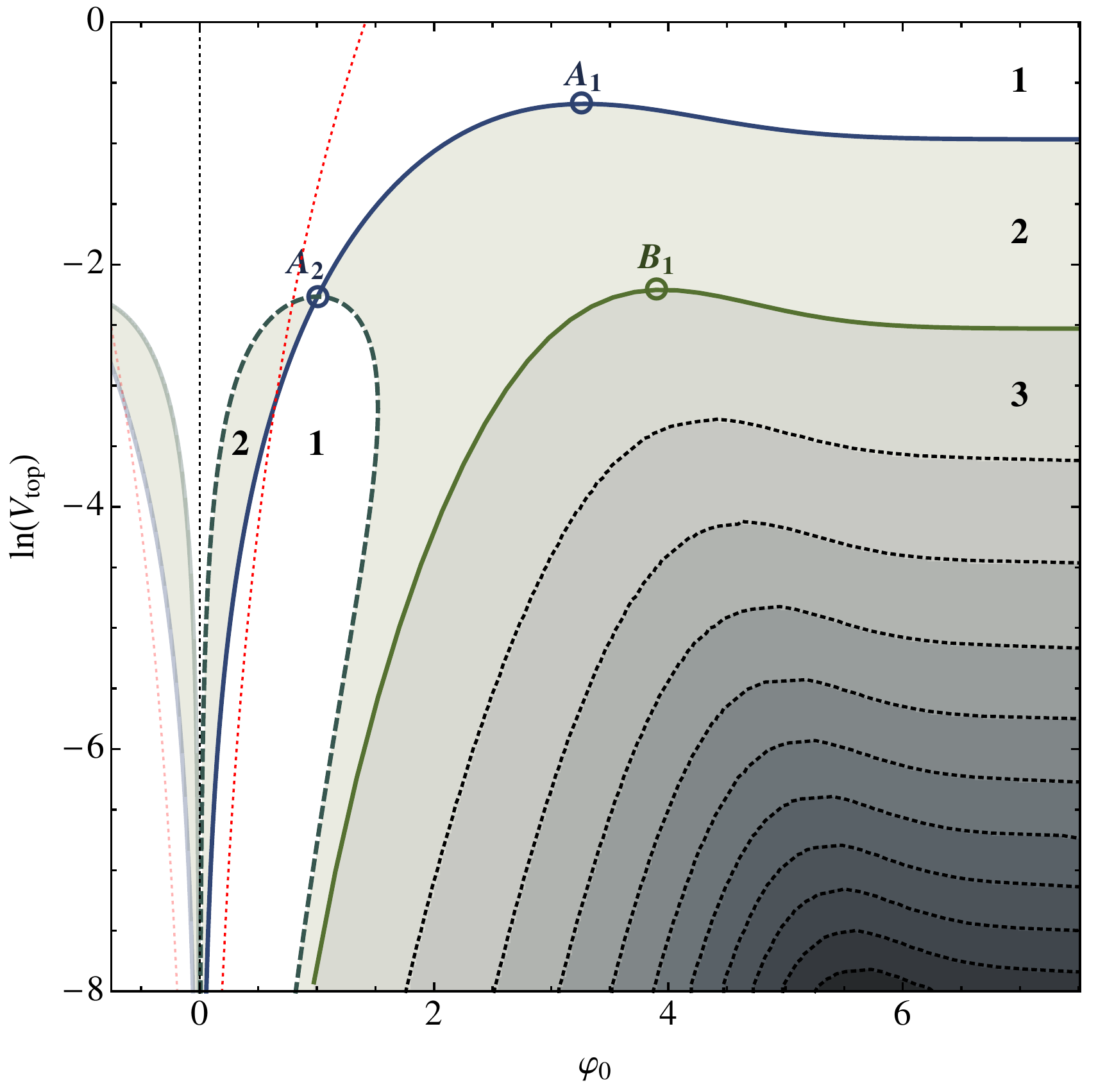}
\end{minipage}%
\begin{minipage}{\smallWidthRight}\flushleft
\includegraphics[width=\smallWidthRight]{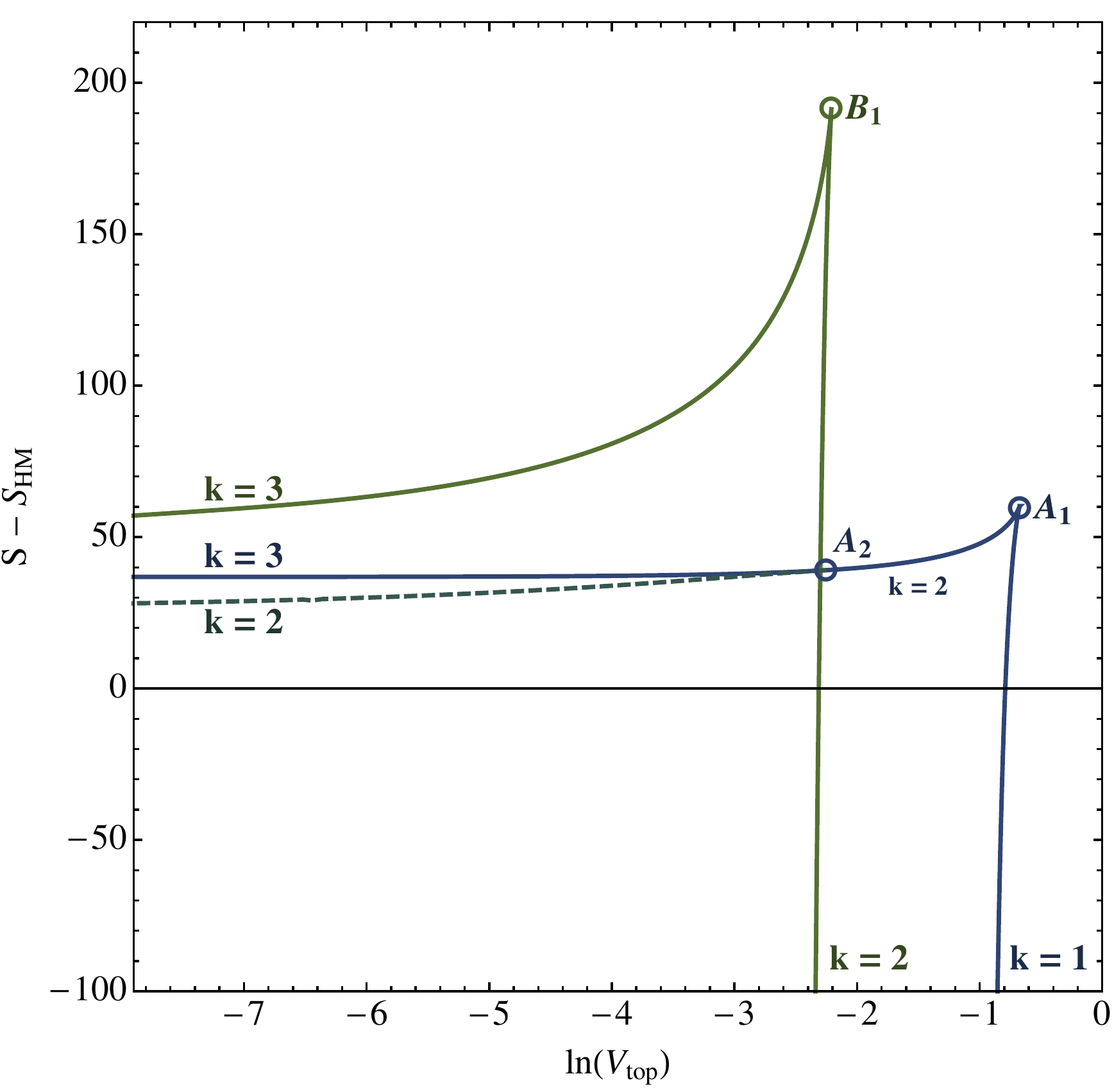}
\end{minipage}%
\caption{\small \label{fig:mapPhi4AndEuclideanActionPhi4}
Left panel: instanton diagram for the potential \eqref{eq:potentialFubini},
with the choice of units $ \lambda = \kappa = 1$. The vertical dotted line represents $\phizero = 0$ and the continuous family of HM solutions. The curved dotted line represents $V = 0$. The asymmetric solutions mentioned in the text correspond to the dashed line emanating from the bifurcation point $A_2;$ these solutions interpolate between two values $\pm \varphi_1, \mp \varphi_2,$ where $\varphi_1, \varphi_2$ are the intersections of a horizontal line and the dashed line. Right panel: difference between euclidean action of the $n=1$ and $n=2$
solutions and the corresponding HM solutions; the number of negative modes, $k$, is also indicated.}
\end{figure}

It is easy to prove that Fubini instantons have one negative mode in their spectrum of linear
perturbations (such solutions are also commonly referred to as \textit{bounces}). Indeed, all Fubini instantons possess a regular zero mode $f_b$ corresponding to an infinitesimal displacement
along the family of solutions,
\begin{equation}
f_b = \partial _{b} \varphi _{b} = \sqrt{ \frac{8}{ \lambda}} \frac{ \eta ^2 - b ^2}{( \eta ^2 + b ^2) ^2} \;.
\end{equation}
Each such zero mode has a single node at $ \eta = b$ and hence, according to nodal theorems, there must exist one node-free solution with a lower eigenvalue -- in other words,
each Fubini instanton possesses a single negative mode.

When gravity is included, this picture changes radically: conformal invariance is lost and at most a finite number of
instantons is left \cite{Lee2013}. Via the rescalings
\begin{eqnarray} \label{eq:resc4one}
g_{\mu\nu}  \equiv  \frac{ \kappa}{ \lambda} \tilde{g}_{\mu\nu} \;, \quad
\varphi & \equiv & \kappa ^{-1/2} \tilde{ \varphi} \;,
\end{eqnarray}
one can set $ \lambda = \kappa = 1$. The rescaling produces a multiplicative factor in front of the euclidean action
\begin{equation} \label{eq:actiontransf1}
S _{ \kappa, \lambda,\Vtop}[g_{\mu\nu}, \varphi] =
\lambda ^{-1} S_{1, 1,\bar{V}_{top}}[ \tilde{g}_{\mu\nu}, \tilde{ \varphi}],
\quad \bar{V}_{top} = \frac{ \kappa ^2 \Vtop}{ \lambda} \;,
\end{equation}
which does not modify any conclusion regarding the existence of instanton solutions, their relative contribution to the tunneling rate
or the spectrum of their negative modes. From now on, we will drop bars and denote by $\Vtop$ the reduced parameter appearing
in the action when $ \lambda = \kappa = 1$.

Assuming $ \Vtop > 0$, one can easily show that the general $O(4)$--invariant  solution is compact. Indeed, in a non--compact solution
the scalar field generally approaches a stationary point of the potential. As the only stationary point for the potential
\eqref{eq:potentialFubini} is located at $V>0$, the corresponding asymptotic geometry cannot be non--compact. The instanton
diagram obtained with this potential is plotted in Figure \ref{fig:mapPhi4AndEuclideanActionPhi4}. Because of the symmetry
of the potential the diagram is also symmetric under $\phizero \rightarrow - \phizero$. The manifest differences with
respect to standard case of Figure \ref{fig:exampleMapAndExampleEuclideanAction} can already be partially explained in terms of
the expected behavior of $n(\phizero)$ at the two extrema of field space:
\begin{itemize}
\item $\phizero = \phitop + \epsilon$: the flatness of the potential near $\phitop$ implies $\ntop = 1$. For this reason, all the
instanton curves bend and do not cross the vertical
axis. This corresponds to the fact that small
perturbations of the scalar field around the HM solution $\varphi = 0$ always ``overshoot'' without oscillating. Indeed, the time--scale
for the potential--induced oscillations to start is roughly
\begin{equation}
\eta_{osc} \sim |V_{, \varphi \varphi}( \phizero)| ^{-1/2} \sim |\varphi_0| ^{-1}
\end{equation}
\begin{figure}[t]
\centering
\includegraphics[width=\fullWidth]{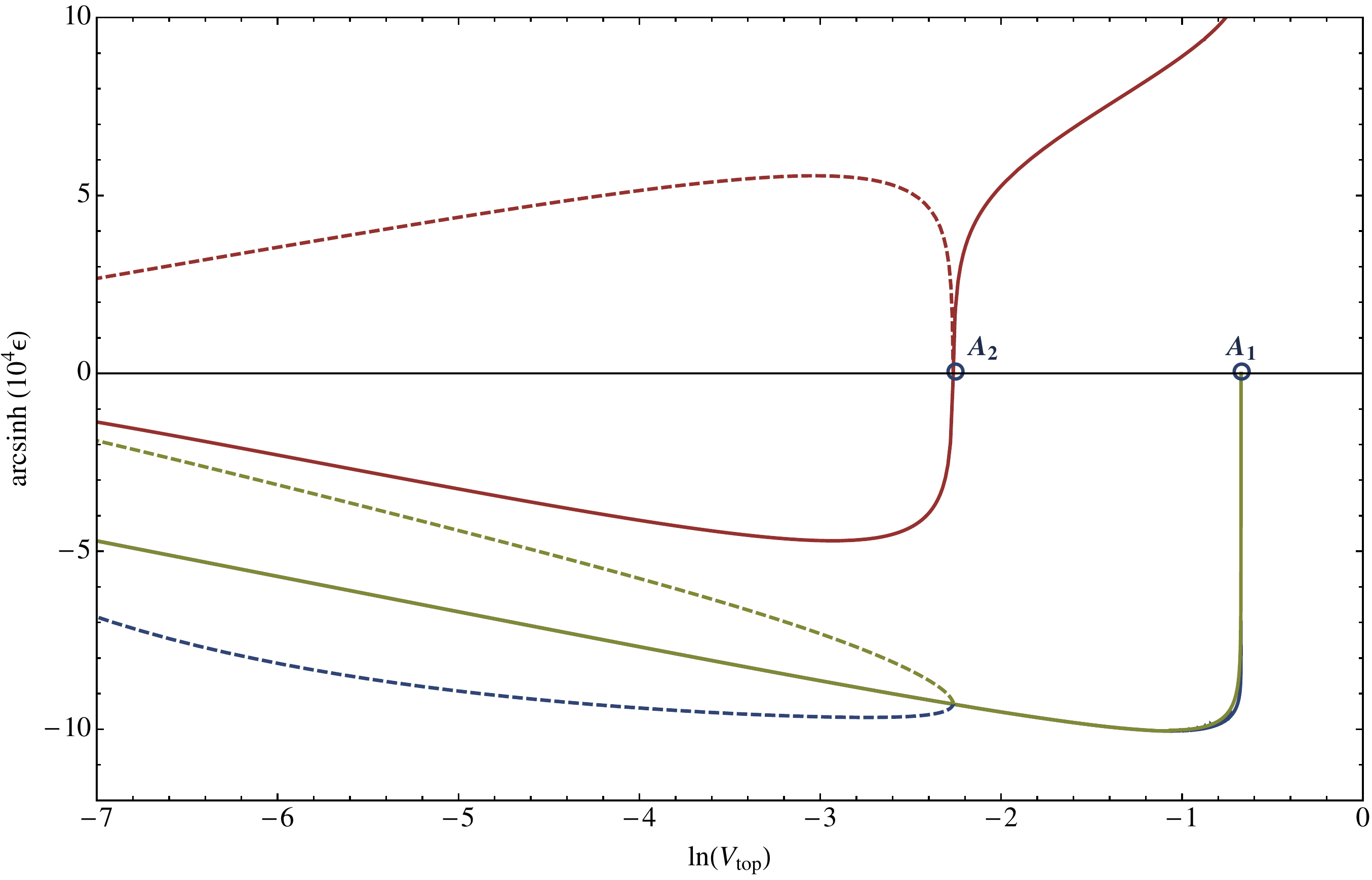} \caption{
\label{fig:negativeModesPhi4}
\small Behavior of the lowest three eigenvalues $\epsilon$ of
the equation \eqref{eq:pertEquation} for $O(4)$--invariant perturbations, for the $n=1$ instantons backgrounds. Dashed lines
correspond to the solutions plotted dashed lines in the left panel of Figure \ref{fig:mapPhi4AndEuclideanActionPhi4}. The two
lowest eigenvalues along the symmetric branch are almost coincident, except near the point $A_1$ where the second eigenvalue
approaches zero.}
\end{figure}
Hence, if $ \phizero$ is made small enough, $ \eta_{osc}$ can be made arbitrarily larger than the HM Hubble radius, and the anti--friction term
makes the scalar field diverge before the oscillations start.
\item $\phizero \rightarrow \phiplus = \infty$: in this limit the overshooting argument clearly does not apply, because the asymptotic
geometry is generally that of euclidean anti--de Sitter space, hence we generally find $ n( \phizero) > 1$ for large values of $\phizero$.
Instead, the instanton curves approach horizontal
asymptotes. This means that the shift $\phizero \rightarrow
\phizero + c$ becomes an approximate solution--generating transformation. Indeed, the corresponding solutions consist of two large
patches of slow--roll pseudo--inflationary solutions in the potential $-V \sim \varphi^4$, glued together by a ``wall'' consisting of the scalar field evolution
near $ \varphi = \phitop$.
\end{itemize}
All the solutions represented in Figure \ref{fig:mapPhi4AndEuclideanActionPhi4}, except the ``strange'' $n=1$ solutions
represented by a dashed line, are symmetric under
$\varphi \rightarrow - \varphi$, $ \eta \rightarrow \bar{ \eta} - \eta$, namely:
\begin{eqnarray}
\eta & \rightarrow & \bar{ \eta} - \eta \;,\\
\varphi & \rightarrow & (-1) ^{n} \varphi \;.
\end{eqnarray}
In particular, they connect $\phizero \equiv \varphi( \eta = 0)$ to $ \phibzero \equiv \varphi( \bar{ \eta}) = (-1) ^{n} \phizero$.
The ``strange'' solutions appear from the bifurcation point $A_2$ and connect the values of $\phizero$
on one side of that point to minus the values on the opposite side, see Figure \ref{fig:mapPhi4AndEuclideanActionPhi4}. The appearance of this bifurcation point and the asymmetry
of the corresponding solutions despite the symmetry of the potential (first noticed in \cite{Hackworth2005}) could not be
observed for the $n>1$ branches in the interval of parameters we considered, but may exist at smaller values of $\Vtop$.

All the non--standard solutions located on the left of
the points $A_1$, $B_1, \ldots$ turn out to
have higher euclidean action than the corresponding HM instanton (see the right panel of Figure
\ref{fig:mapPhi4AndEuclideanActionPhi4}). Moreover, focusing on the $n=1$ solutions, we observed that across the
points $A_1$ and $A_2$ additional negative modes appear. In particular, the symmetric branch possesses two negative modes
between $A_1$ and $A_2$, and three negative modes below
$A_2$. This is in full agreement with
our expectations, at these two points the horizontal
slope of an instanton curve signals the
presence of a zero mode: the instantons located at $A_1$ and $A_2$ are critical solutions. On the other hand, the asymmetric
$n=1$ branch possesses two negative modes, and has lower euclidean action than the symmetric one (see the right panel
of Figure \ref{fig:mapPhi4AndEuclideanActionPhi4}). For this reason,  the
additional negative mode of the latter can then be interpreted as the perturbation generating the transition from the symmetric
to the non--symmetric branch.

This qualitative analysis is confirmed by an explicit computation, along the different $n=1$ branches, of the three lowest eigenvalues
of the perturbation equation \eqref{eq:pertEquation} (see Figure \ref{fig:negativeModesPhi4}). Across $A_1$ and $A_2$, a positive
eigenvalues becomes negative on the symmetric branch.\footnote{Note that on the right of
$A_1$, the computation of the negative modes according to \eqref{eq:pertEquation} becomes inconsistent, as the
$\mathcal{Q}$ function appearing in the perturbation potential becomes negative in a finite range of values of $ \eta$. How to consistently work out the perturbation modes in this case is currently an open problem.}
These results show that the number of negative modes can be different from the number of oscillations of an instanton
solution. Furthermore, several solutions with the same number of negative modes can co--exist: e.g. the $n=1$ symmetric
solutions below
$A_2$ and the $n=2$ solutions on the left of $B_1$, both possessing three negative modes.

The solutions located to the right of
the critical points $A_1$, $B_1, \ldots$ appear as standard CdL solutions, for which the number of negative modes $k$ coincides with
the number of oscillations $n$. However, as the potential \eqref{eq:potentialFubini} is unbounded from below, they
only exist in small ranges of values of $\Vtop$. Moreover, as shown in Figure \ref{fig:mapPhi4AndEuclideanActionPhi4},
the scalar field takes values for which the potential is negative for all these solutions, which  are therefore not
directly relevant for tunneling from de Sitter to de Sitter/Minkowski space.

At this point, it is worth stressing that the structure of the space of solutions would be considerably harder to understand without the use of instanton diagrams. As we explained above, the standard method consists in choosing a particular theory (in this case the value of $\Vtop$) and looking for the values of $\phizero$ across which $n(\phizero)$ is discontinuous. This corresponds to analyzing the behavior of $n(\phizero)$ on horizontal slices of the plot in left panel of Figure \ref{fig:mapPhi4AndEuclideanActionPhi4}. It is evident that the pattern of discontinuities depends in a non--trivial way on the slice one chooses, i.e. on the particular value of $\Vtop$ under consideration, so that the appearance and properties of different solutions may look entirely uncorrelated.

\begin{figure}[t!]
\begin{minipage}{\thirdWidthLeftDelayed}
\flushleft
\includegraphics[width=\thirdWidthLeftDelayedInner]{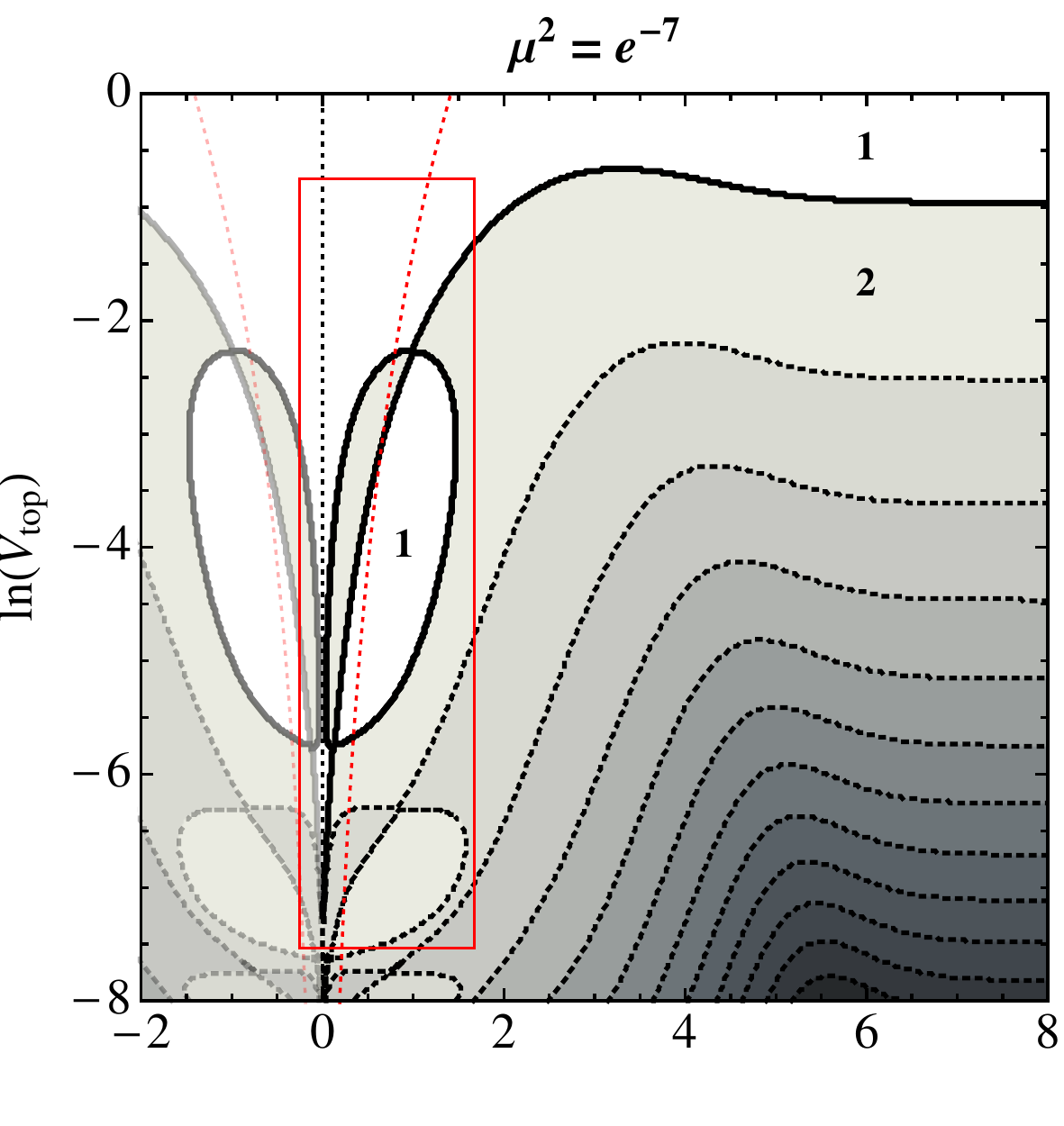}
\end{minipage}%
\begin{minipage}{\thirdWidthRightDelayed}
\flushleft
\includegraphics[width=\thirdWidthRightDelayedInner]{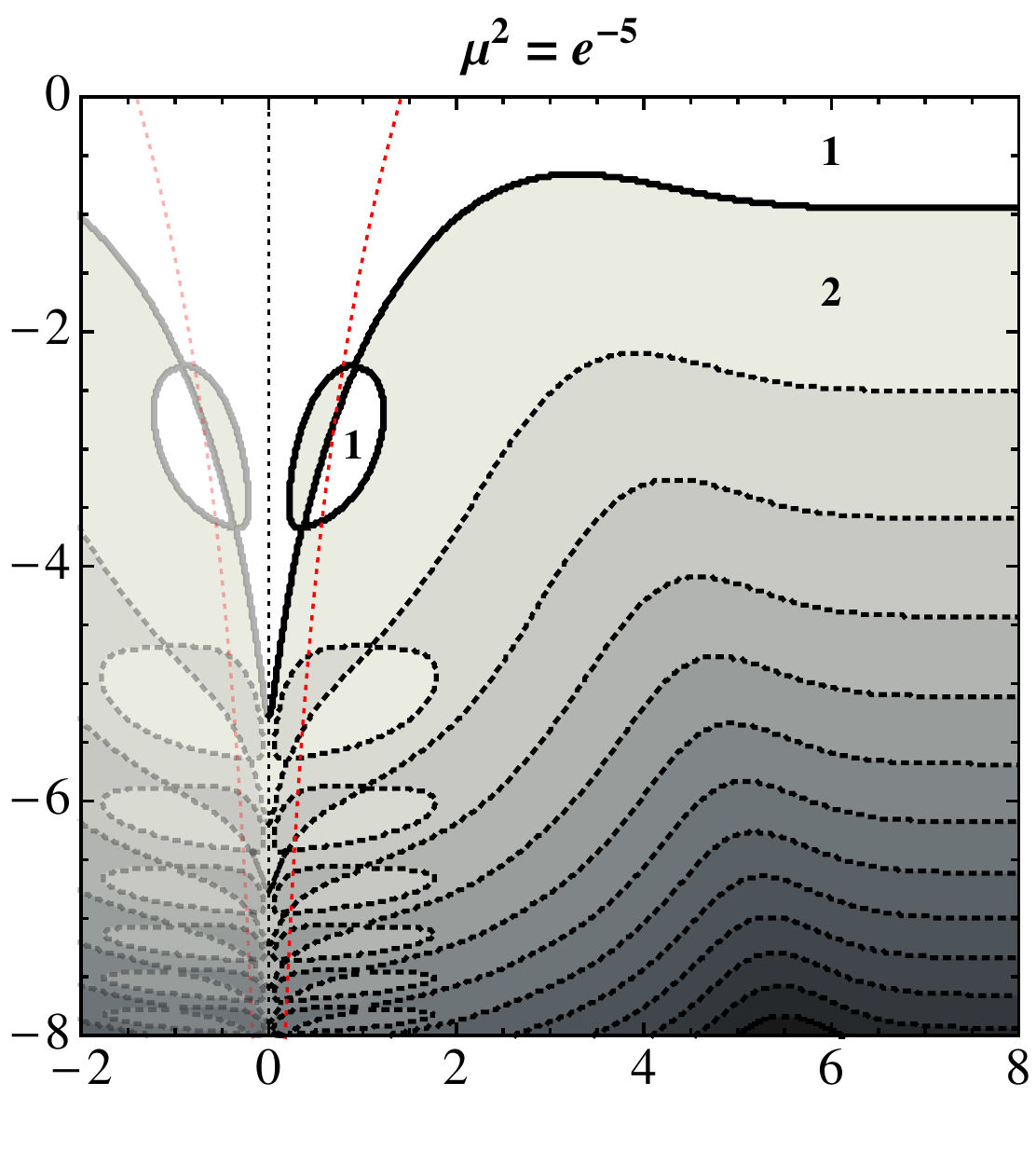}
\end{minipage}%
\begin{minipage}{\thirdWidthRightDelayedInner}
\flushleft
\includegraphics[width=\thirdWidthRightDelayedInner]{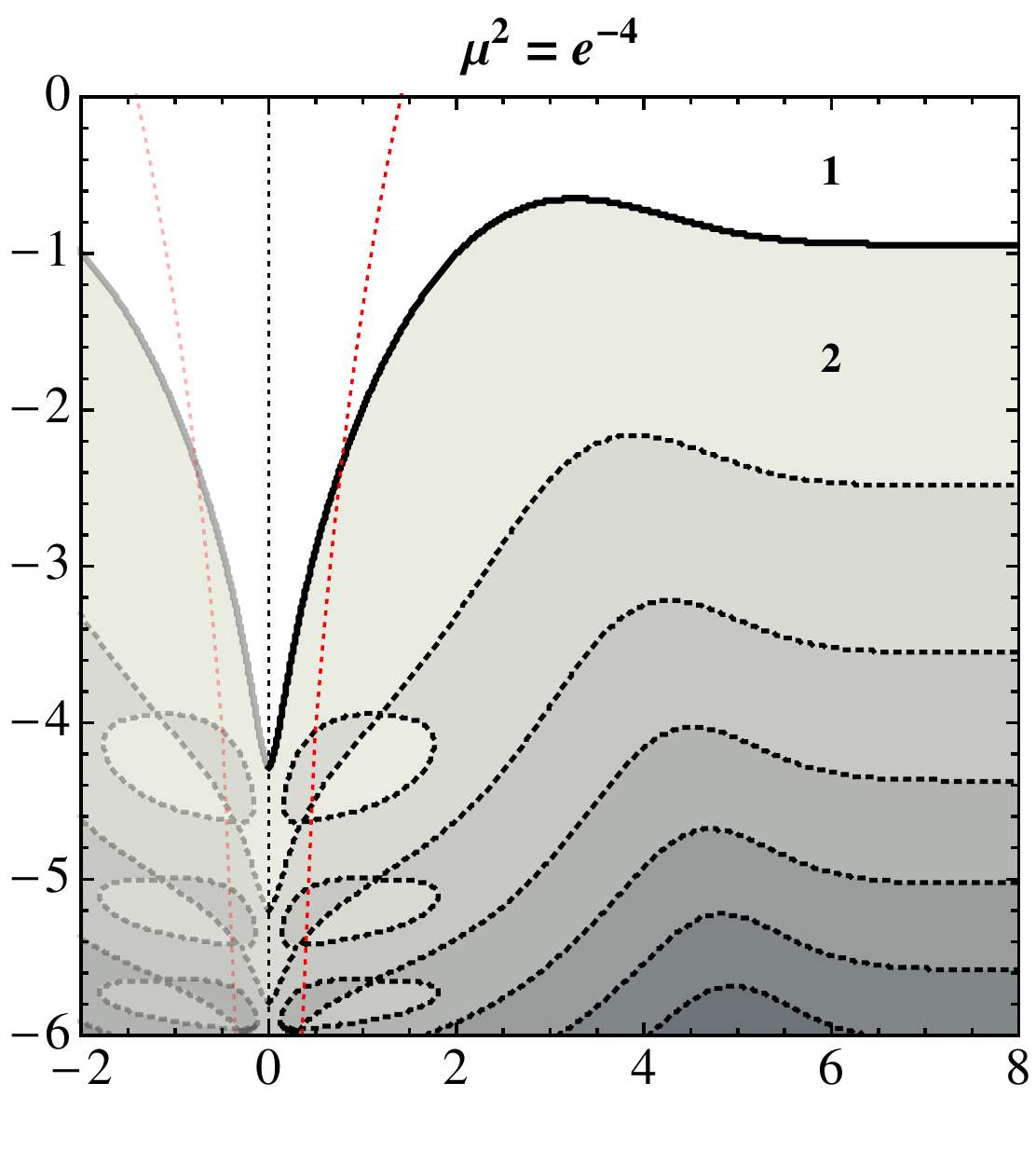}
\end{minipage}\\
\begin{minipage}{\thirdWidthLeftDelayed}
\flushleft
\includegraphics[width=\thirdWidthLeftDelayedInner]{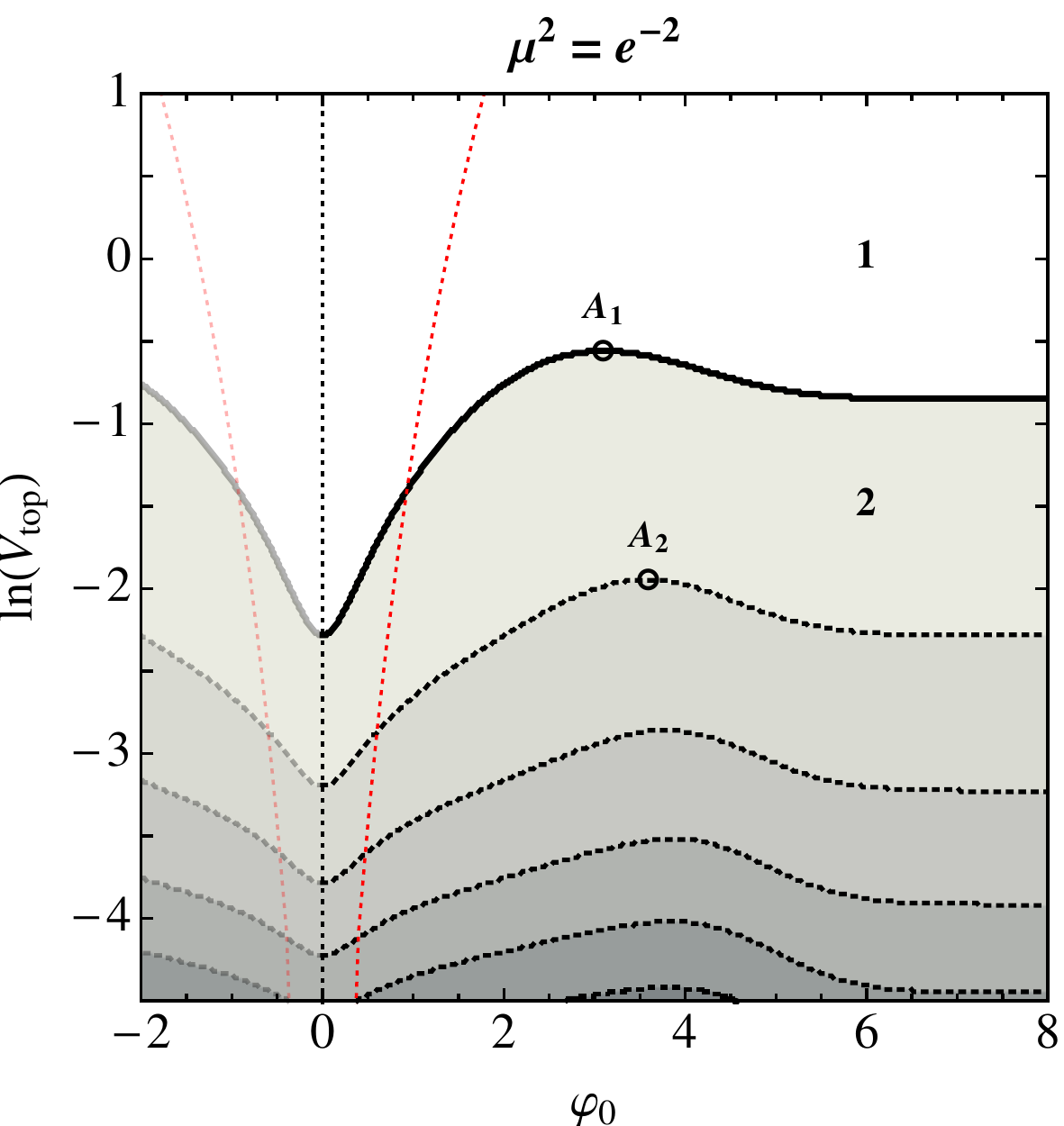}
\end{minipage}%
\begin{minipage}{\thirdWidthRightDelayed}
\flushleft
\includegraphics[width=\thirdWidthRightDelayedInner]{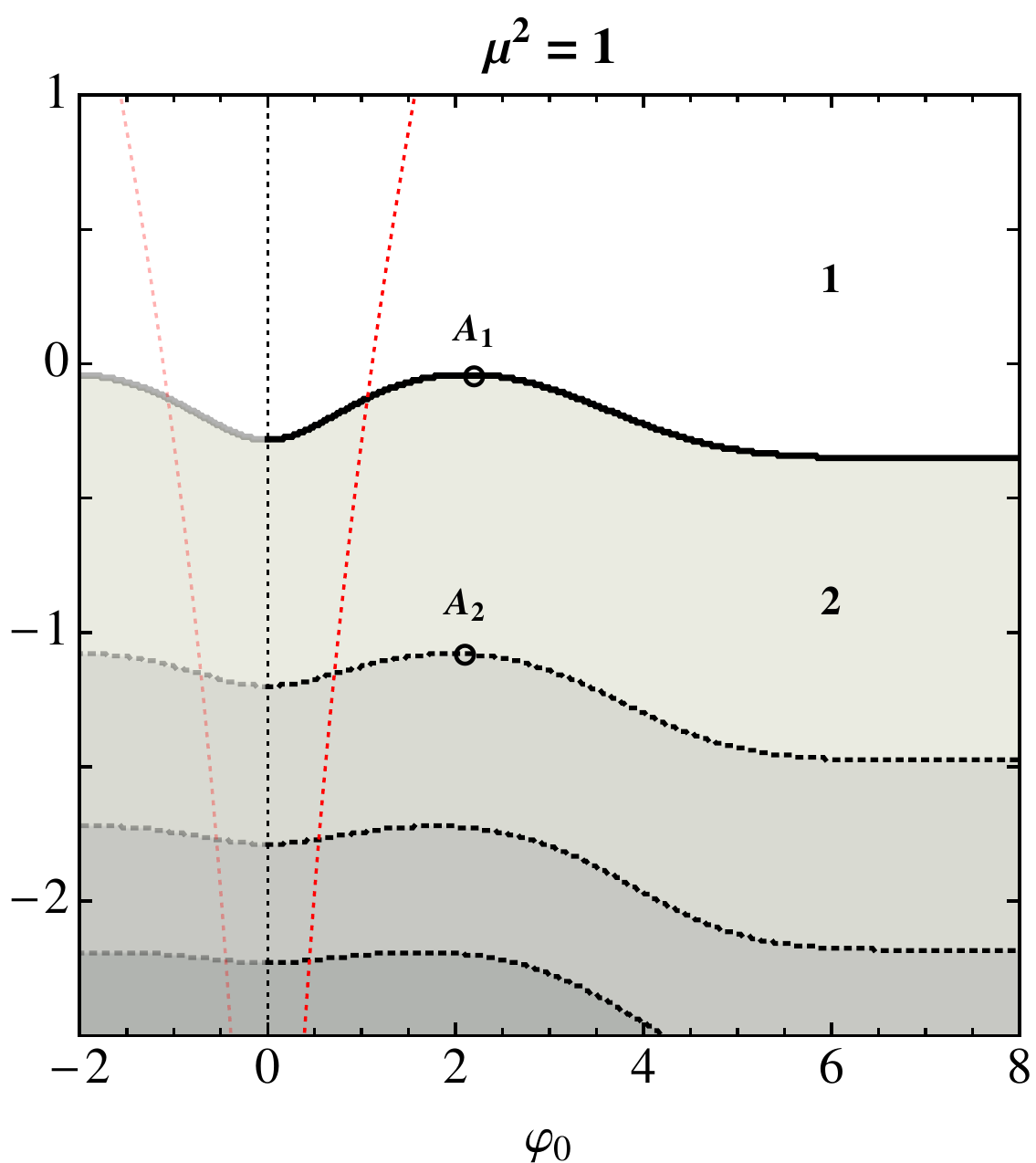}
\end{minipage}%
\begin{minipage}{\thirdWidthRightDelayedInner}
\flushleft
\includegraphics[width=\thirdWidthRightDelayedInner]{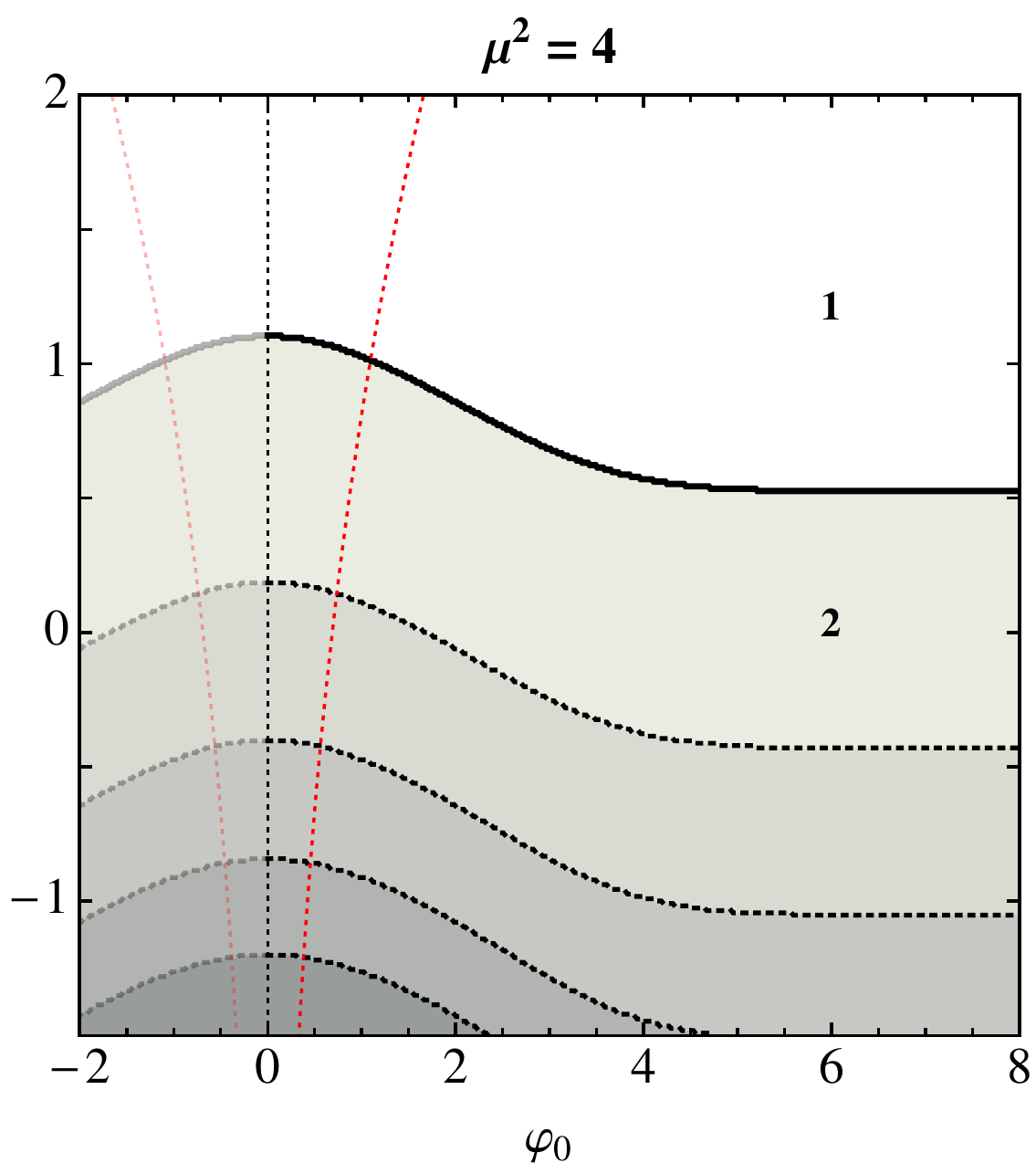}
\end{minipage}\caption{\label{fig:zoologyMasses}Instanton diagrams for various values of the reduced mass parameter $ \mu ^2$ in the potential
\eqref{eq:potential24reduced}. The dotted lines represent $V(\phizero) = 0$ as in Figure \ref{fig:mapPhi4AndEuclideanActionPhi4}. The region enclosed in the box in the top, leftmost panel is depicted in the left panel of Figure
\ref{fig:detailMassTerm}. }
\end{figure}

\section{Intermediate case: $V = \Vtop - \frac{ \mu ^2}{2} \varphi ^2 - \frac{1}{4} \varphi ^4$ \label{sect:massTerm}}
In the previous section we have seen that the flatness of the potential at $\phitop$ allows the existence of critical
instantons with a non--trivial scalar field and, correspondingly, the presence of  new instanton branches with additional
negative modes. It is then natural to ask how this picture evolves to the standard one when the curvature of the potential
is increased from zero.

To address this question, we consider the potential
\begin{equation} \label{eq:potential24}
V = \Vtop - \frac{ \mut ^2}{2} \varphi ^2 - \frac{ \lambda}{4} \varphi ^4 \;.
\end{equation}
As in the previous case, by an appropriate choice of units we may set $ \lambda = \kappa = 1$,
\begin{equation} \label{eq:potential24reduced}
V = \bar{V}_{top} - \frac{ \bar{ \mut} ^2}{2} \varphi ^2 - \frac{1}{4} \varphi ^4 \;.
\end{equation}
The rescaled parameters appearing in \eqref{eq:potential24} are related to the original parameters by
\begin{eqnarray}
\bar{V}_{top} & = & \frac{ \kappa ^2 \Vtop}{ \lambda} \;,\\
\bar{ \mu} ^2 & = &  \frac{ \kappa \mu ^2}{ \lambda} \;.
\end{eqnarray}
Using the same argument as for the $ \mu ^2=0$ case, one can easily prove that all $O(4)$--invariant solutions of the field
equations with the potential \eqref{eq:potential24} are compact.

In Figure \ref{fig:zoologyMasses} we present several instanton diagrams corresponding to different values of $ \mu ^2$.
The presence of a mass term allows $n(\phitop+ \epsilon)$ to take values greater than one for sufficiently small $\Vtop$,
in agreement with \eqref{eq:bounddS}. In the instanton diagrams, this is related to the fact that the instanton curves
now join the $ \varphi = 0$ axis, and thus they are continuously related to the HM critical solutions, like in the standard case depicted in Figure
\ref{fig:exampleMapAndExampleEuclideanAction}.

\begin{figure}[t]
\begin{minipage}{\thirdWidthLeft}
\flushleft
\includegraphics[width=\thirdWidthRight]{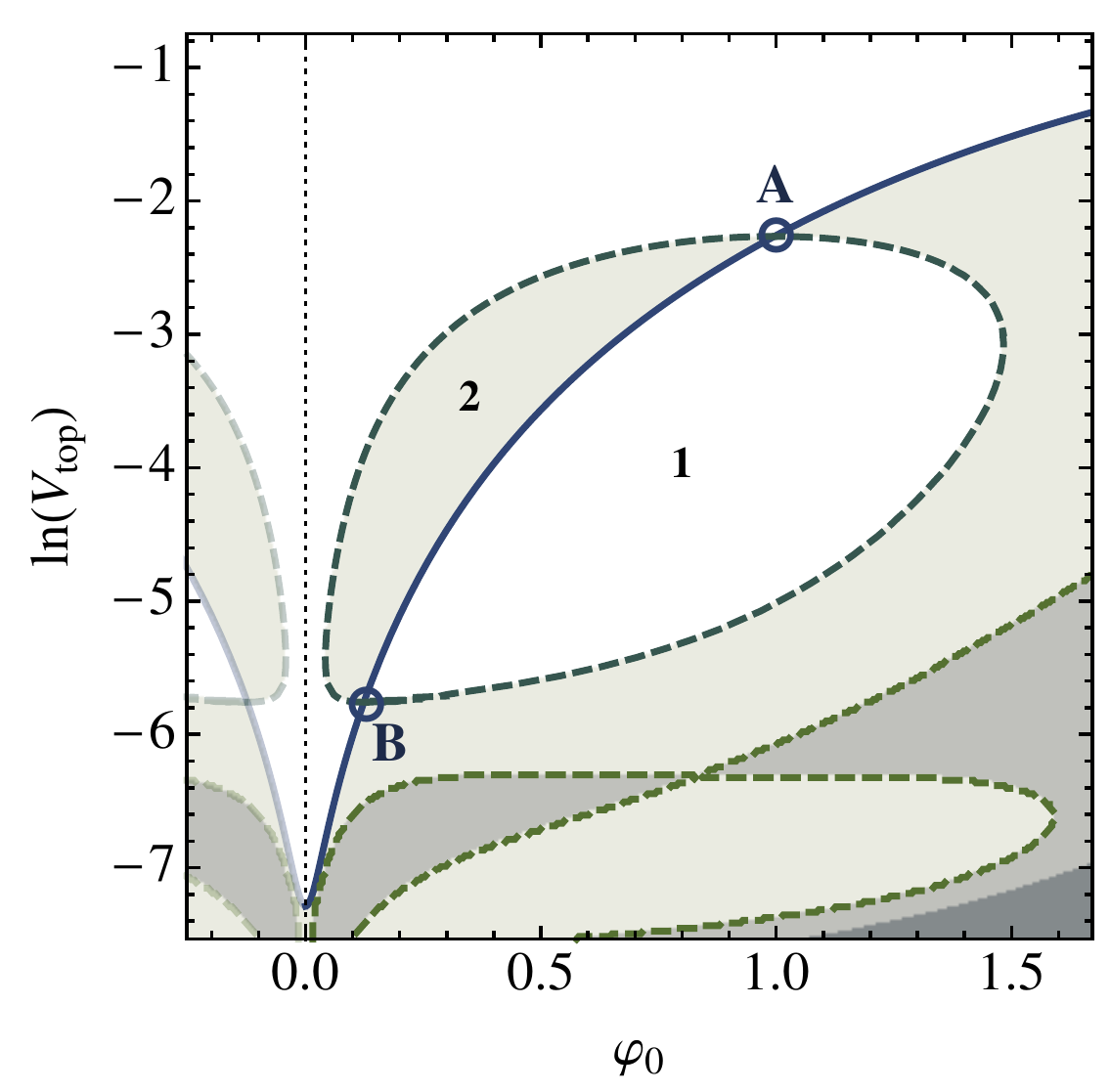}
\end{minipage}%
\begin{minipage}{\thirdWidthLeft}
\flushleft
\includegraphics[width=\thirdWidthRight]{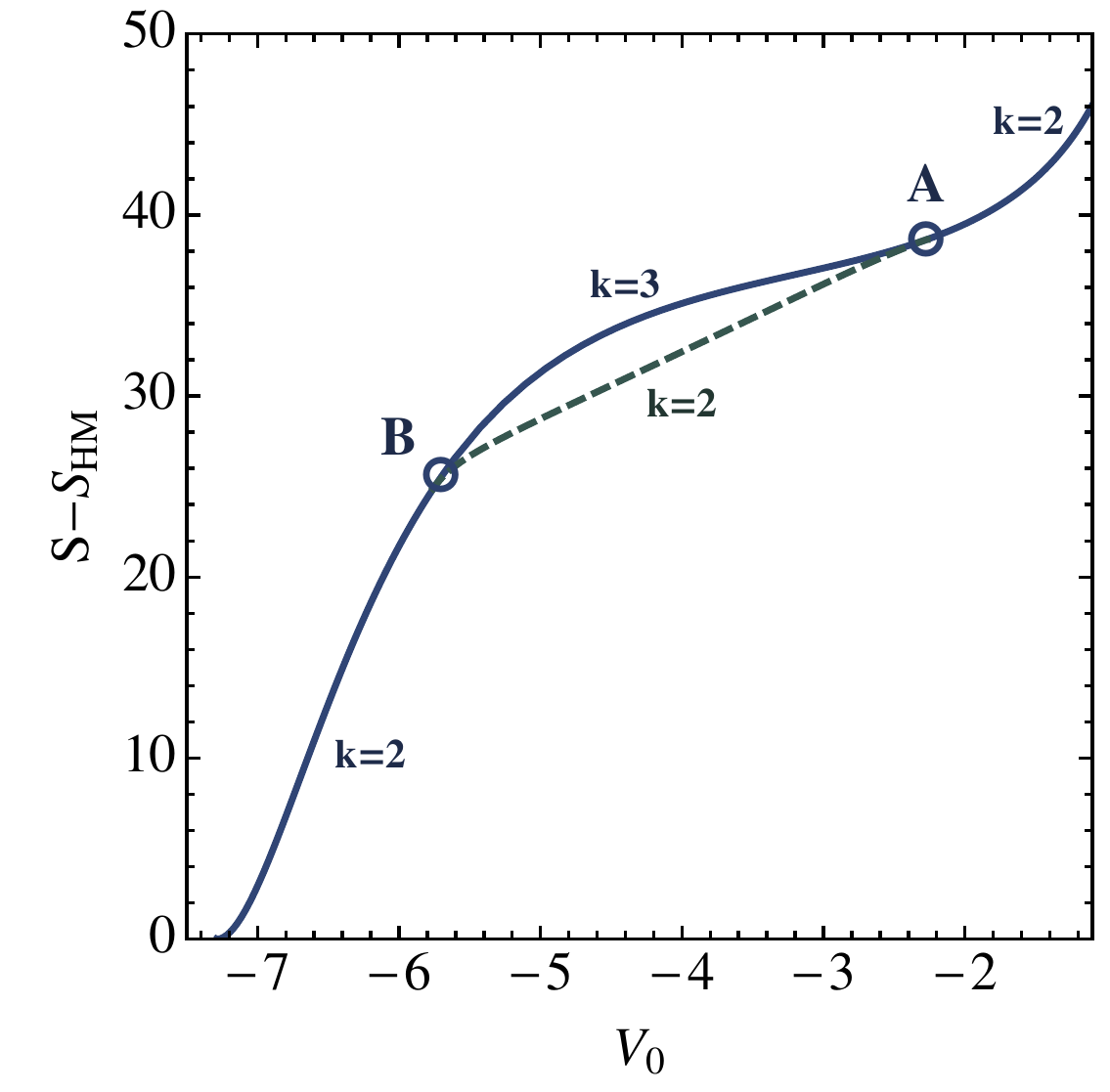}
\end{minipage}%
\begin{minipage}{\thirdWidthRight}
\flushleft
\includegraphics[width=\thirdWidthRight]{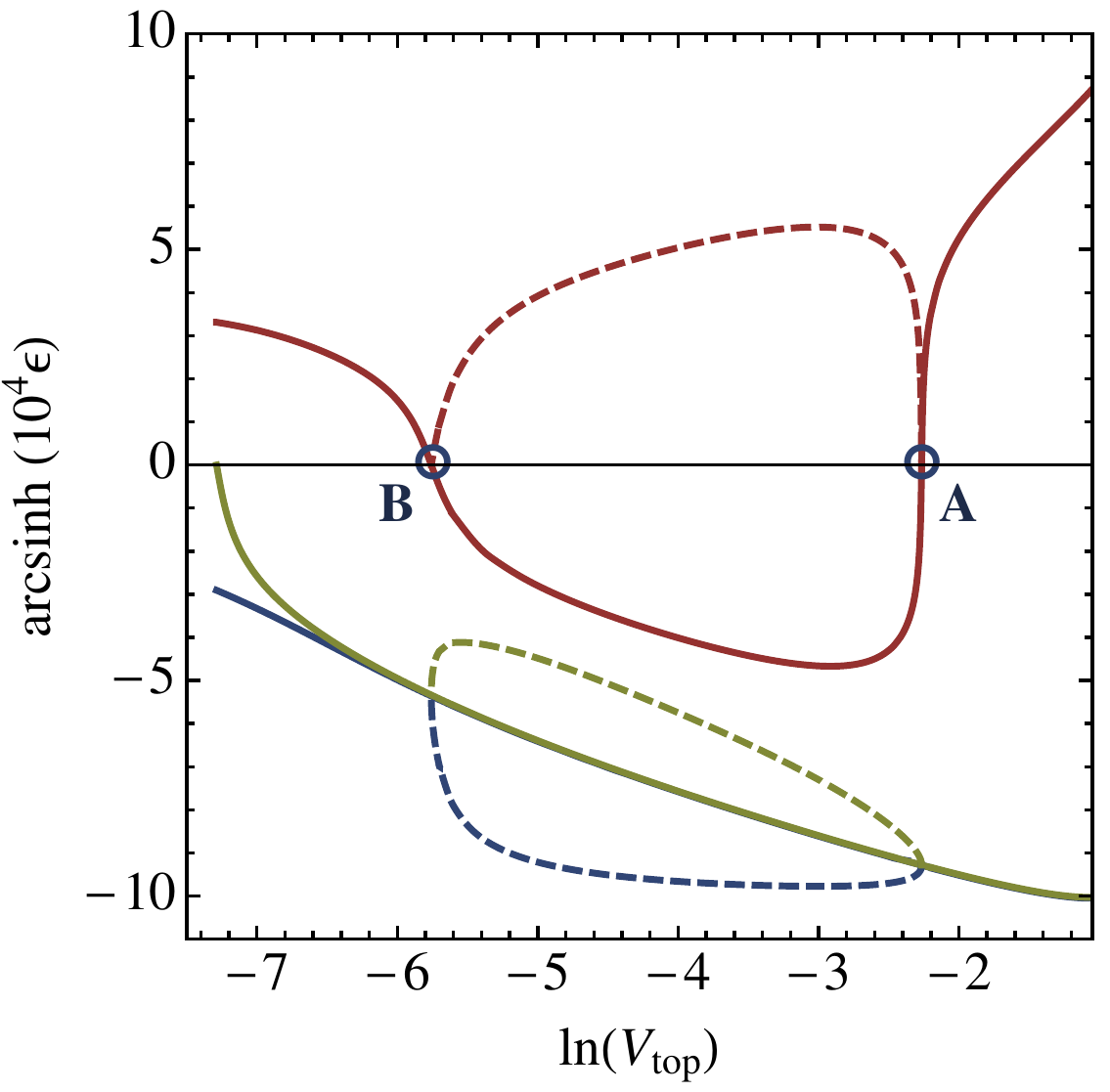}
\end{minipage}%
\caption{ \label{fig:detailMassTerm}Left panel: detail of the instanton diagram for $ \mu ^2 = e^{-7}$ (see Figure \ref{fig:zoologyMasses}).
Center panel: difference between the euclidean action of the $n=1$ instantons and the corresponding HM solutions across the
critical points $A$ and $B$. Right panel: three lowest eigenvalues of the $n=1$ instanton solutions; dashed lines correspond to the
non--symmetric $n=1$ solutions depicted in dashed lines on the left panel.}
\end{figure}

For small values of $ \mut ^2$, the bifurcation corresponding to the critical point along the $n=1$ curve is now
associated to a second bifurcation at smaller values of $\Vtop$, giving rise to a ring--like structure
(see also Figure \ref{fig:detailMassTerm}). The same structure appears on all the $n>1$ branches we were able to explore,
suggesting that a single bifurcation should also be present, at values of $\Vtop$ below our numerical limitations,
in the $ \mut ^2 = 0$ case. In going through the ring in the direction of decreasing
$\Vtop$, the symmetric branches acquire an additional negative mode and, correspondingly, their euclidean action is greater than that of the
corresponding non--symmetric branches (Figure \ref{fig:detailMassTerm}, center panel). The eigenvalue associated to this
mode turns positive again across the new bifurcation point, close to the critical HM solution
(Figure \ref{fig:detailMassTerm}, right panel). For these small values of $ \mut ^2$, the properties of the solutions
located at larger $ \phizero$ remain the same as in the $\mut ^2 = 0$ case, discussed in the previous section. Moreover,
also like in the massless case, the non--standard solutions appearing on the left
of the critical points $A_1, A_2, \ldots$ always have greater euclidean action than the corresponding HM instantons.

When $\mut ^2$ is increased, the ring--like structures gradually shrink and disappear, starting from the
one located on the $n=1$ branch. Moreover, at the critical values $ \mut_{c,1}, \mut_{c,2},\ldots$, the critical instantons located at
$A_1,A_2, \ldots$ approach the $\phizero = 0$ axis and ``disappear'' by joining the critical HM solutions. For this reason,
when  $ \mu > \mut_{c,n} $ the non--standard branches are no longer present and only ``standard'' $n$--oscillating instantons
with $k=n$ negative modes survive: $\ntop > 1$ is again a necessary condition for the existence of solutions.
In this regime, except for the different behavior at large $\phizero$ due to the  unboundedness of the potential,
the instanton diagrams are qualitatively similar to the one presented in Figure \ref{fig:exampleMapAndExampleEuclideanAction}.

The critical values $ \mut _{n} ^{c}$ can be determined by analyzing the perturbations of the critical HM solutions. Indeed,
as one can clearly see in Figure \ref{fig:zoologyMasses}, a small perturbation of the $n$--critical HM instanton will
result in a singular $n+1$ (resp. $n$) solution if $ \mu < \mu _{c,n}$ (resp. $ \mu > \mu _{c,n}$). More precisely,
the behavior of the curve of $n$--oscillating instantons at small $(\phizero-\phitop)$ is determined by the number of oscillations of the
solutions:
\begin{eqnarray} \label{eq:expansionSolution}
\varphi & = & \varphi( \eta, \phizero) \;,\\  \label{eq:expansionSolution2}
\rho & = & \rho( \eta, \phizero) \;,
\end{eqnarray}
in the theory admitting the $n$--critical HM solution
\begin{equation}
n(n+3) = \frac{ 3 \mu ^2}{ \kappa V_n} \;.
\end{equation}
For small $\phizero$, the solution (\ref{eq:expansionSolution},\ref{eq:expansionSolution2})
can be written as an expansion in the value of $(\phizero-\phitop)$:
\begin{eqnarray}
\varphi( \eta, \phizero) & = & \phitop + (\phizero- \phitop)\, \varphi_{(1)}( \eta) + (\phizero- \phitop) ^2 \, \varphi_{(2)}( \eta) + \ldots \;,\\
\rho( \eta, \phizero) & = & \rho_{HM} + (\phizero- \phitop)\, \rho_{(1)}( \eta) + (\phizero- \phitop) ^2\,  \rho_{(2)}( \eta) + \ldots \;.
\end{eqnarray}
The functional coefficients $( \varphi_{(k)}, \rho_{(k)})_{k \geq 1}$ are the $k$--th order $O(4)$--invariant perturbations of the background critical
 HM solution:
\begin{equation}
\rho_{HM} = H_n\, \sin{(H _{n} \eta)}, \quad H_n = \left( \frac{ \mu ^2}{n(n+3)} \right) ^{1/2} \;.
\end{equation}
The boundary conditions specifying these perturbation modes follow from those for the full solution:
\begin{eqnarray}
\varphi( \eta = 0, \phizero)  =  \phizero &\quad \Longrightarrow  \quad &
\left\{ \begin{array}{rcl}
\varphi_{(1)}(0) & = & 1 \;,\\
\varphi_{(k)}(0) & = & 0, \quad k > 1 \;,
\end{array}
\right. \\
\rho( \eta = 0, \phizero)  =  0 & \quad \Longrightarrow \quad & \rho_{(k)}(0) = 0 \;, \\
\rho'( \eta = 0, \phizero)  =  1 & \quad \Longrightarrow \quad & \rho _{(k)}'(0) = 0 \;.
\end{eqnarray}

\noindent
For general values of $\phizero$, we expect the solution \eqref{eq:expansionSolution} to be singular. This singularity
is reflected in the singular behavior of the perturbation modes $ \varphi_{(n)}$ near the south pole of the HM instanton
$ \bar{ \eta} = \pi / H_n$. Even though the unbounded growth of the perturbation modes signals the breakup of perturbation
theory, the behavior of the lowest order singular mode is likely to determine the singular behavior of the full solution.
In the case of $n=1$ solutions, the third mode $ \varphi_{(3)}$ is the lowest order singular mode \cite{Tanaka1992}:
\begin{eqnarray}
\varphi_{(3)} & \stackrel{ \eta \rightarrow \bar{ \eta} }{ \propto} &- \frac{32 + \nu ^2 + 18 \lambda}{( \bar{ \eta} - \eta) ^2} \;, \\
\nu & \equiv & \frac{2V ^{(3)}_{top}}{  \kappa ^{1/2} \mu ^2} \;,\\
\lambda & \equiv & \frac{ 2 V ^{(4)}_{top}}{ 3\kappa  \mut ^2}  \;.
\end{eqnarray}
The transition between the overshooting ($n=1$) and undershooting ($n=2$) behaviors corresponds to change in the sign of
the divergence of $ \varphi_{(3)}$:
\begin{equation} \label{eq:bound1}
8\kappa\, \mu ^{4} _{c,1} +  3V ^{(4)} _{top}\, \mu ^2 _{c,1}  + (V ^{(3)}_{top} ) ^2 = 0 \;.
\end{equation}
In our case $ \kappa = 1$, $ V ^{(4)} _{top} = -6$ and $ V ^{(3)} = 0$:
\begin{equation} \label{eq:bound1b}
\mu _{c,1}  ^{2} = \frac{9}{4} \;.
\end{equation}
This result can be stated differently, saying that when the fourth derivative of the potential $\phitop$ is more negative
than the critical value:
\begin{equation}
V ^{(4)} _{c} = - \frac{8 \kappa \mu _{c,1} ^2}{ 3} - \frac{  (V ^{(3)}_{top} ) ^2}{ 3 \mu _{c,1} ^2} \;,
\end{equation}
new $n=1$ instanton solutions appear in a class of theories with $\ntop \lesssim 1$.  The critical value \eqref{eq:bound1}
can be derived in a more rigorous way by studying the properties of the action functional in the vicinity of the critical
HM solution \cite{Balek2005}. Using similar techniques, one can prove analytically that $n=1$
istantons appearing at small $ ( \phizero - \phitop)$ when $ \mu ^2 \lesssim \mu_{c,1} ^2$  always have
greater euclidean action than the corresponding HM solutions \cite{Demetrian2007}, which we numerically
verified (see Figure \ref{fig:detailMassTerm}, center panel).

In conclusion, when the curvature $ \mu ^2$ of the potential at its top is nonzero, the small--$ ( \phizero - \phitop)$
solutions can be studied as small perturbations of the critical HM solution. However, focusing on the $n=1$ solutions,
when $ \mu ^2 \ll \mu_{c,1}$, the position of the critical point $A_1$ remains approximately fixed (see Figure \ref{fig:zoologyMasses})
and, over a wide range of values for $\Vtop$, non--standard $n=1$ solutions  are found which cannot be described as perturbations
of HM. The existence of the ring--like structure and the non--symmetric branch therein are a striking manifestation of this fact. This
suggests, in agreement with the expectation expressed in \cite{Hackworth2005}, that the existence of these non--standard branches
cannot be directly related to the shape of the potential at $ \varphi = \phitop$, but is rather determined by the full
structure of the potential.

\section{More realistic flat potentials \label{sect:realistic}}
\begin{figure}[t]
\begin{minipage}{\thirdWidthLeftDelayed}
\flushleft
\includegraphics[width=\thirdWidthLeftDelayedInner]{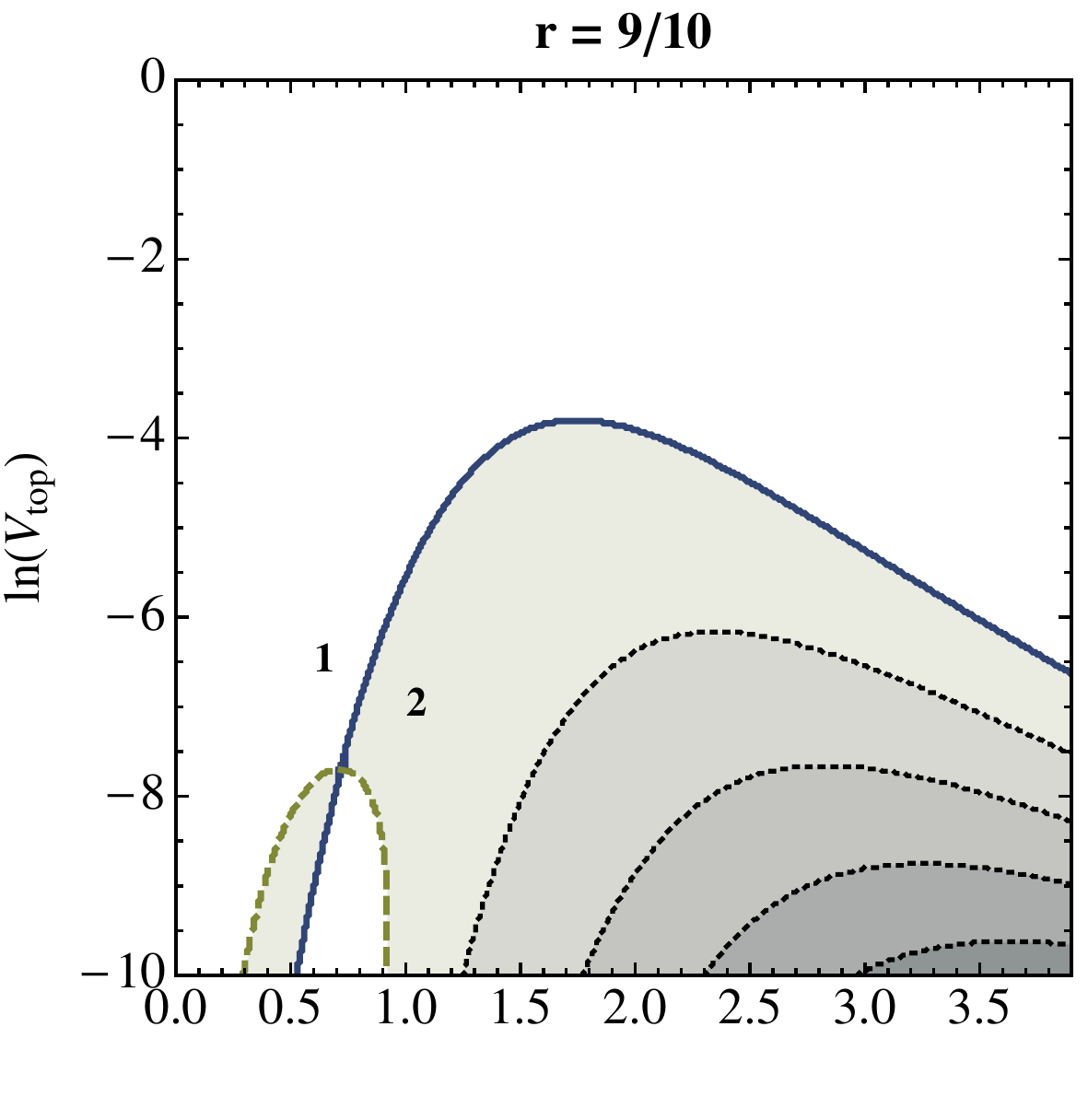}
\end{minipage}%
\begin{minipage}{\thirdWidthRightDelayed}
\flushleft
\includegraphics[width=\thirdWidthRightDelayedInner]{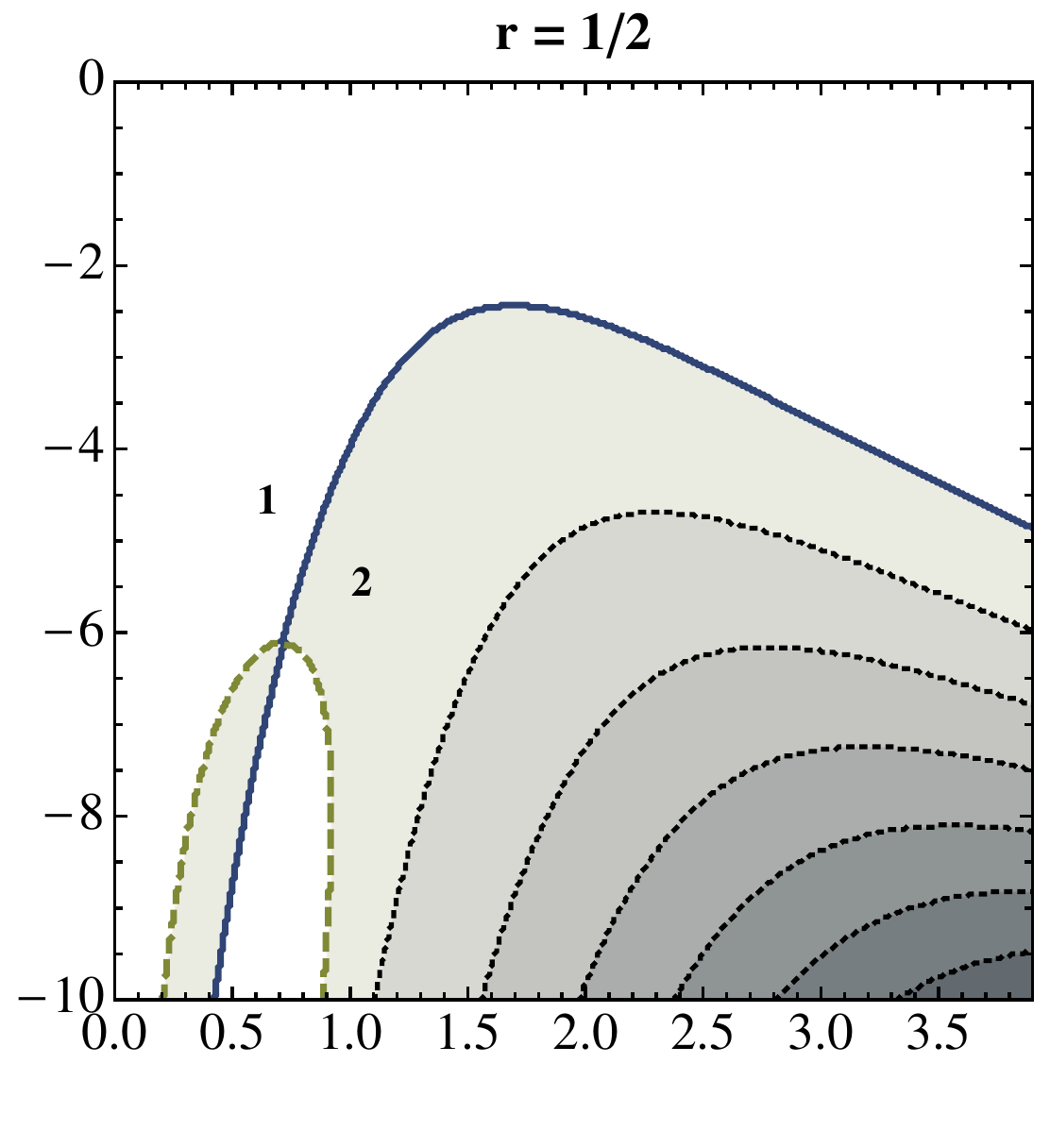}
\end{minipage}%
\begin{minipage}{\thirdWidthRightDelayedInner}
\flushleft
\includegraphics[width=\thirdWidthRightDelayedInner]{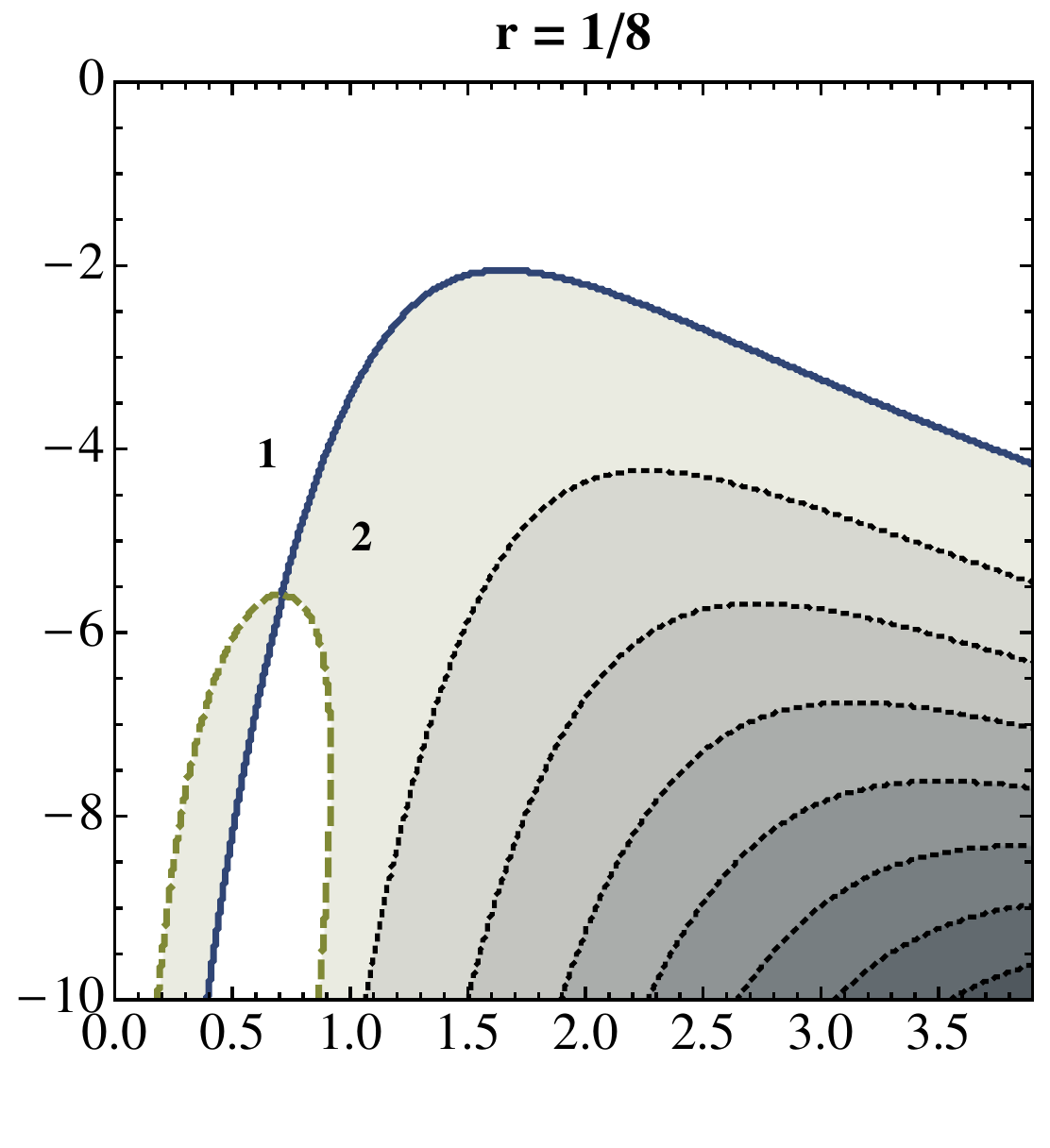}
\end{minipage}\\
\begin{minipage}{\thirdWidthLeftDelayed}
\flushleft
\includegraphics[width=\thirdWidthLeftDelayedInner]{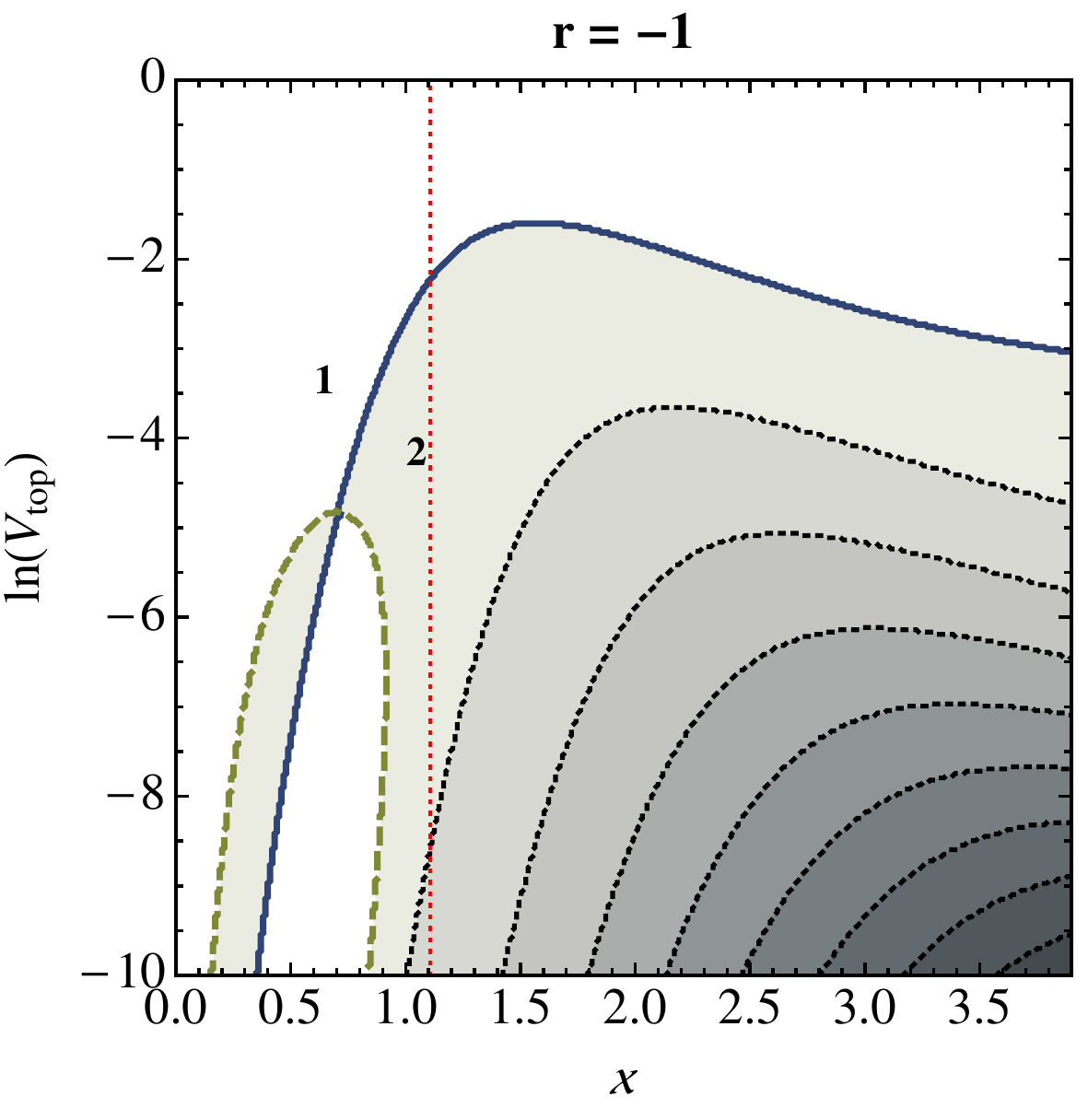}
\end{minipage}%
\begin{minipage}{\thirdWidthRightDelayed}
\flushleft
\includegraphics[width=\thirdWidthRightDelayedInner]{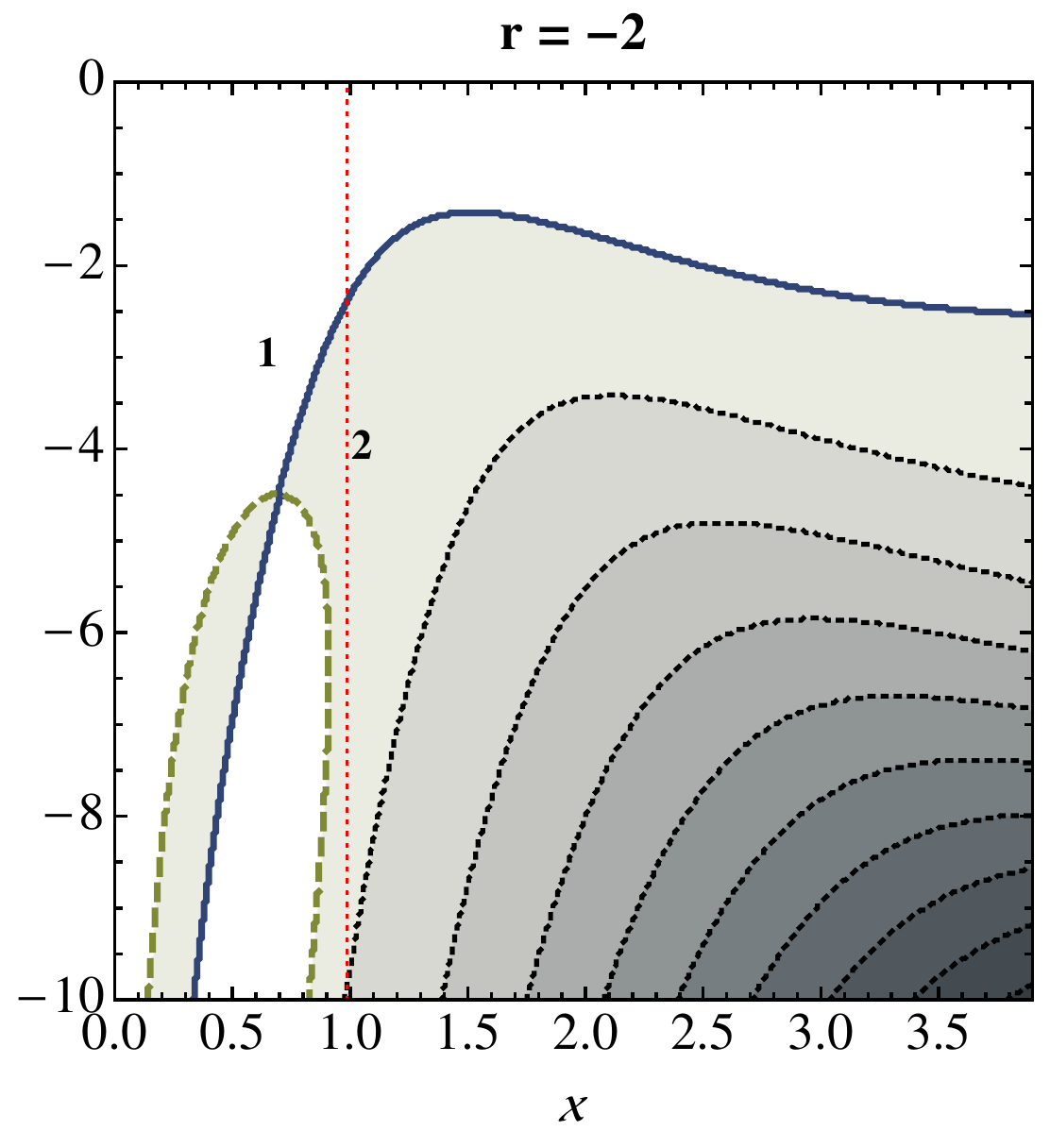}
\end{minipage}%
\begin{minipage}{\thirdWidthRightDelayedInner}
\flushleft
\includegraphics[width=\thirdWidthRightDelayedInner]{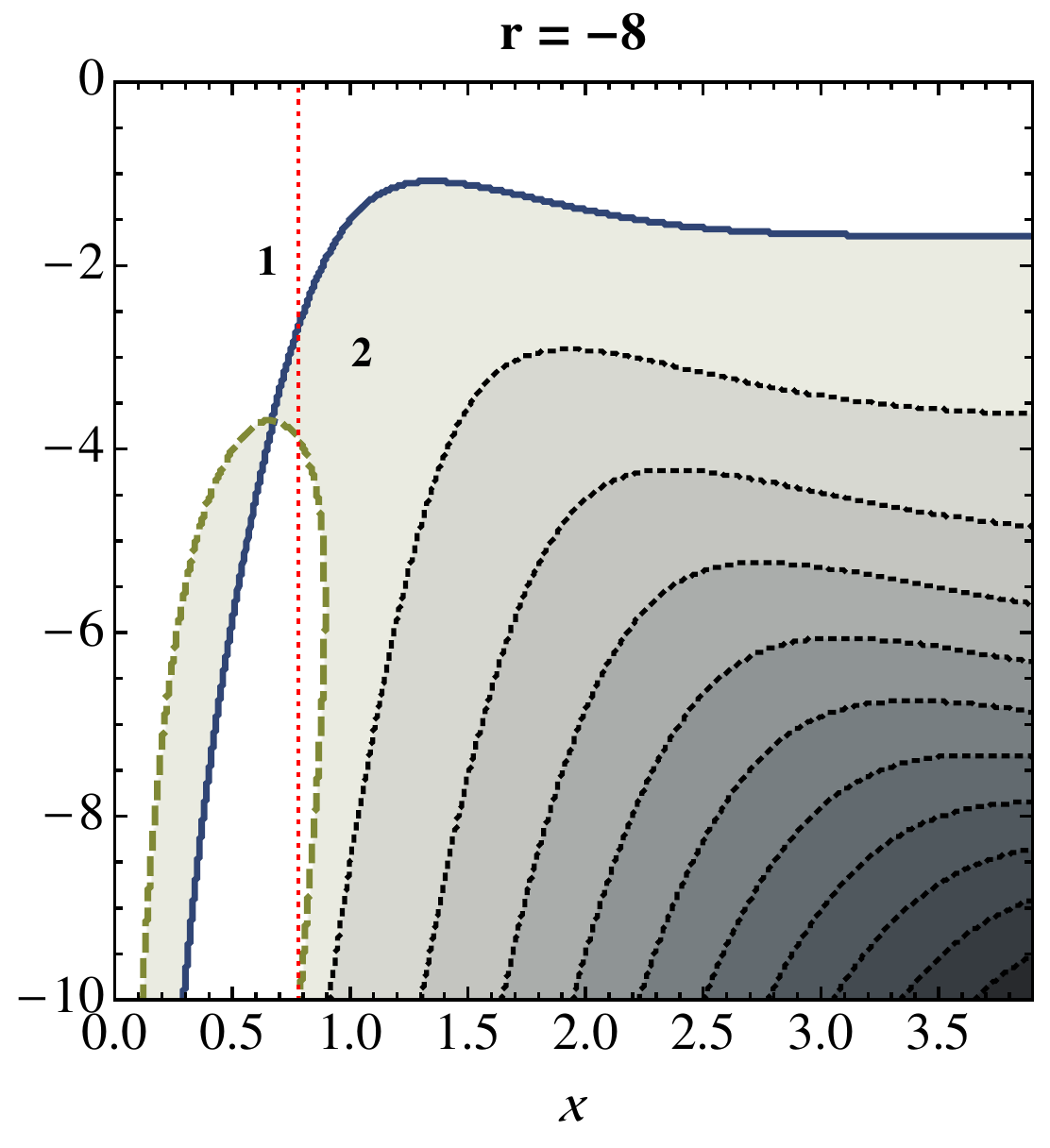}
\end{minipage}%
\caption{\label{fig:zoologyRegularizations} Instanton diagrams for the potential \eqref{eq:potential46} and different
values of the ratio \eqref{eq:ratior}. The dotted lines represent $V = 0$ as in the previous figures.}
\end{figure}

\begin{figure}[t]
\begin{minipage}{\smallWidthLeft}
\flushleft
\includegraphics[width=\smallWidthRight]{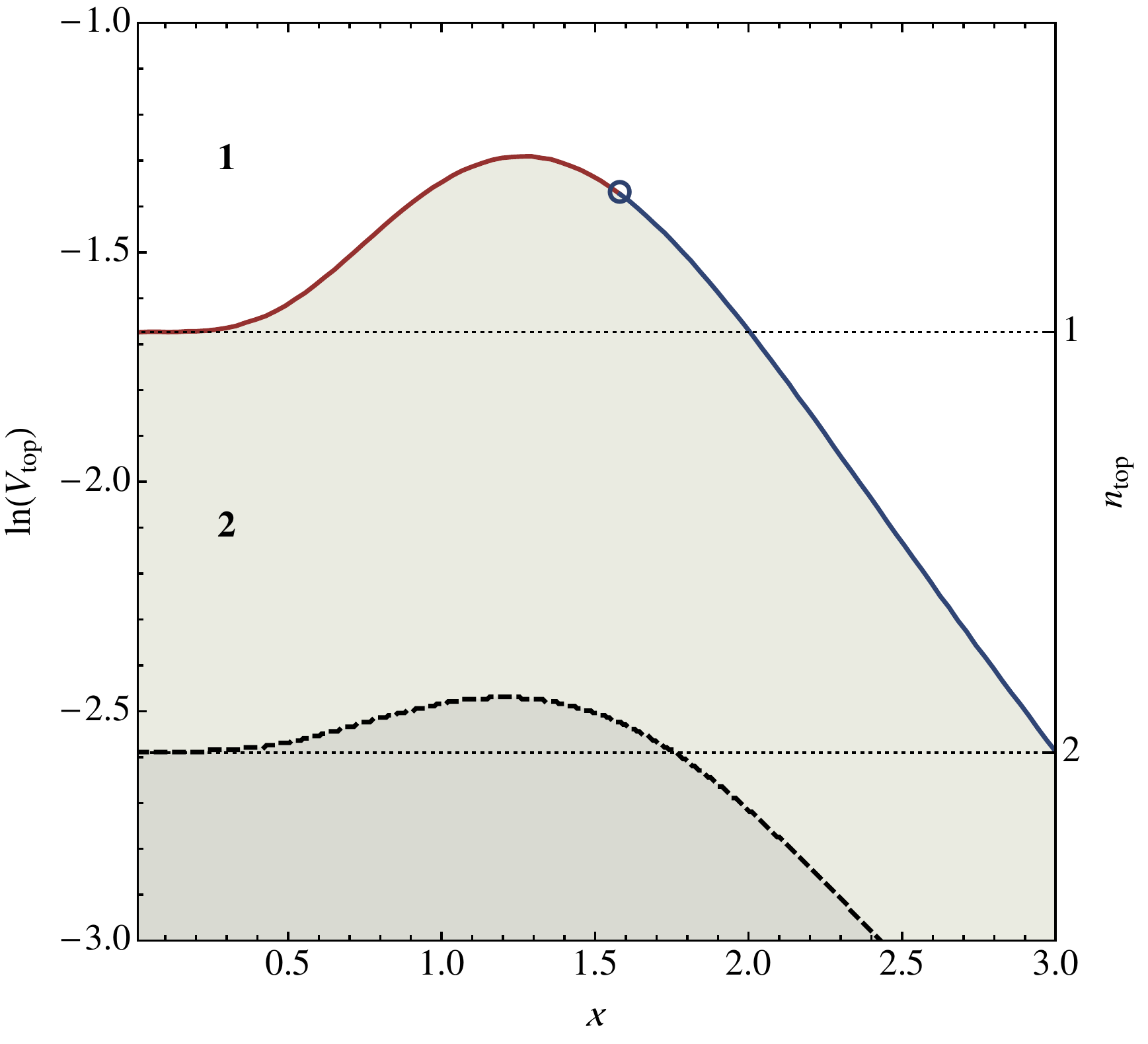}
\end{minipage}%
\begin{minipage}{\smallWidthLeft}
\flushleft
\includegraphics[width=\smallWidthRight]{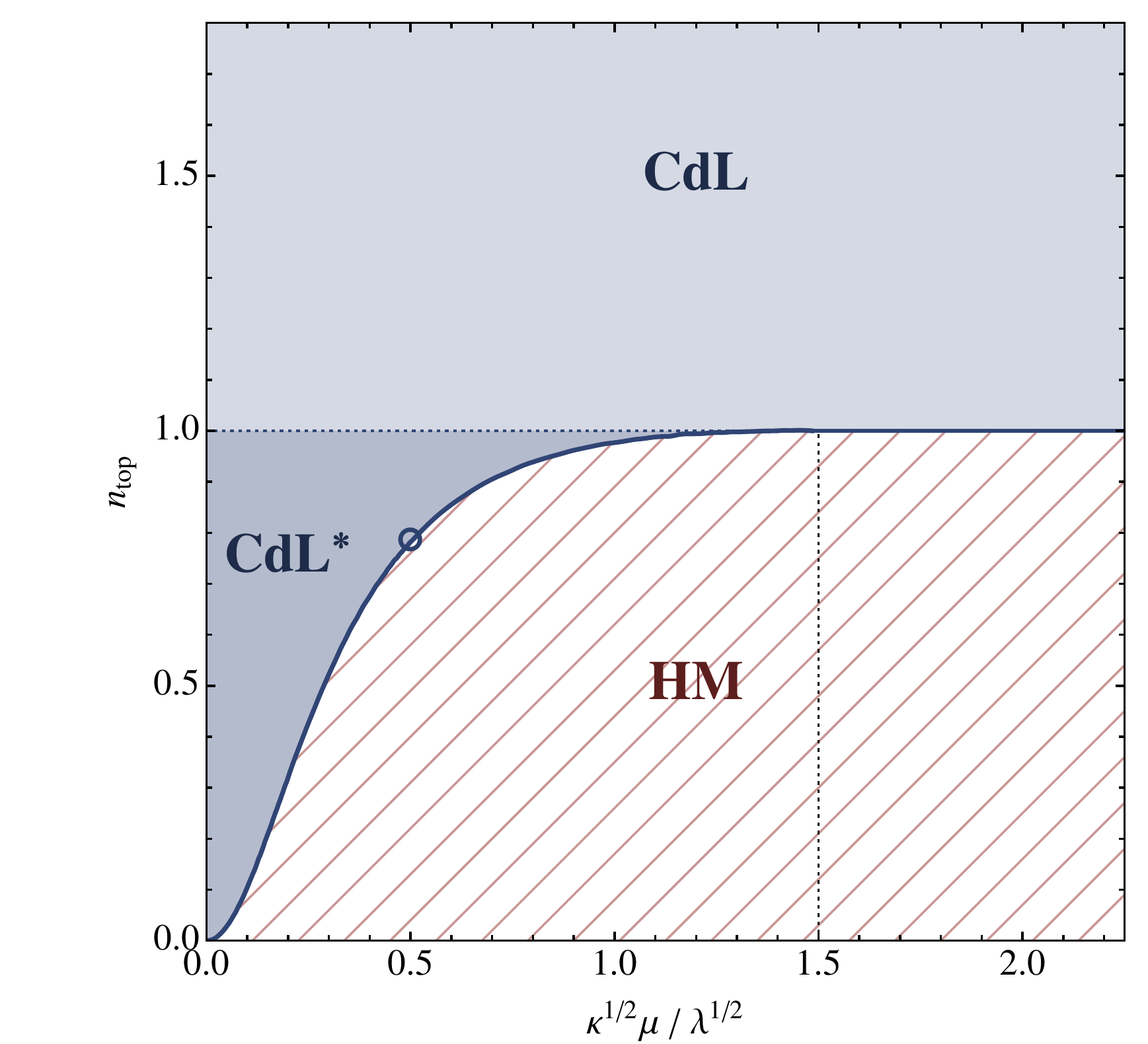}
\end{minipage}%
\caption{\label{fig:phaseDiagram} Left panel: instanton diagram for the theory \eqref{eq:potential246} with $r = 1/4$,
$ \mu = 0.5$ and the choice of units $ \lambda = \kappa = 1$. The part of the curve on the right of the highlighted point
represents $n=1$ solutions with more negative euclidean action than the HM solution at equal $\Vtop$. Right panel:
phase diagram for the dominant tunneling channel for the potential \eqref{eq:potential246} with $r = 1/4$. For
$ \mu < \mu_{c,1} = 3/2$ \eqref{eq:bound1b},  $n=1$ instantons with $\ntop < 1$ persist which dominate the decay rate. The
highlighted point corresponds to the solution depicted on the left panel.}
\end{figure}

\subsection{Positive potentials}
In the previous two sections, we studied the existence of instanton solutions connecting vacua separated by
 flat and nearly flat potential barriers. However, as illustrated in Figures \ref{fig:mapPhi4AndEuclideanActionPhi4} and
 \ref{fig:zoologyMasses}, most of these solutions extend to ranges of field space in which $V < 0$. In order to clarify the
 role of these negative values of $V$ in the existence of non--standard $\ntop<1$ solutions, we considered a regularized
 version of the scalar potential \eqref{eq:potentialFubini}:
 \begin{equation} \label{eq:potential46}
 V = \Vtop - \frac{ \lambda}{4} \varphi ^4 + \frac{t}{6} \varphi ^6 \;.
 \end{equation}
Via the rescalings (\ref{eq:resc4one}) we set again $ \lambda = \kappa = 1$. In order to consider
a class of theories with $V > 0$ and let $\Vtop$ vary, we parametrically fix $t$ so as to keep the ratio between the
vacuum energy densities at the degenerate vacua $\varphi _{ \pm}$ and at $ \phitop$ constant:
\begin{equation} \label{eq:ratior}
r = V_{ \pm}/\Vtop \;.
\end{equation}
When $r<0$, one has $ V _{ \pm} < 0$ and non--compact solutions could exist for which the scalar field approaches $ \phiplus$ or $\phiminus$ asymptotically. However, for such a non--compact solution the quantity $ \frac{1}{2} \varphi ^{ \prime 2} - V$ is decreasing, and its asymptotic value $ V_ {+} = V_{-}$ is necessarily smaller than its initial value $V_0$. In the case of our potential $V \geq V _{\pm}$ and the only non--compact solutions are the euclidean--AdS $ \varphi( \eta) = \varphi _{ \pm}$ solutions.

In Figure \ref{fig:zoologyRegularizations} we present instanton diagrams for several values of $r$. The reduced field variable $x$
already introduced in Section \ref{sect:standardCdL} is used in order to visualize solutions which have $\phizero$ very close to
the true vacuum value $\phiplus$:
\begin{equation}
\phizero \equiv \phiplus \left(  1- e^{- x ^2}  \right) \;.
\end{equation}
The left
part of the diagrams show a clear resemblance with the one describing the $V = \Vtop - \varphi^4/4$ case.
In particular, the bifurcation along the $n=1$ branch remains visible. We also checked
that, by adding a small mass term, the ring structure described in the previous section appear at least along the $n=1$
branch. This suggests that the features of the non--standard branches are not related to negative values of the potential. Indeed,
the appearance of a bifurcation point had first been noticed in the case of a positive potential \cite{Hackworth2005}.
At the same time, as expected from the overshooting argument, when $r>0$ all the instanton curves bend down
as $x$ increases,
in such a way that  $n( \phiplus - \epsilon, \Vtop) = 1$
for any value of $\Vtop$.  In this respect, the large $x$ part of these instanton diagrams
is analogous to the one for the standard CdL case (see Figure \ref{fig:exampleMapAndExampleEuclideanAction}).
On the other hand, when $r < 0$ the curves appear again to approach constant values of $\Vtop$ as in Figure
\ref{fig:mapPhi4AndEuclideanActionPhi4}, which described the $r = - \infty$ case.

Focusing on the $r>0$ case, we stress again that in the presence of a flat potential barrier, for sufficiently small $\Vtop$
non--standard solutions coexist with standard CdL--type solutions with $ \phizero$ very close to $\phiplus$. The existence
of both types of solutions in spite of the flatness of the potential barrier ($\ntop = 0$ in the present case) can be related to
the presence of the critical solution  sitting at the top
of each instanton curve. The euclidean
action and the number of negative modes of these solutions follow the behaviors already described in Section \ref{sect:fubini}.
In particular, as already noticed in \cite{Demetrian2007} for a potential with $\ntop \lesssim 1$,
``standard'' $n=1$ solutions located sufficiently at the right of
the critical points have more negative euclidean action than the corresponding HM solutions. Possessing a single negative mode, these
solutions are the dominant tunneling channel for this class of theories.

As expected from the results of the previous section, this qualitative picture remains valid even if the potential is
not exactly flat at $\varphi = \phitop$, e.g.:
\begin{equation} \label{eq:potential246}
V = \Vtop - \frac{ \mu ^2}{2} - \frac{ \lambda}{4} \varphi ^4 + \frac{t}{6} \varphi ^6 \;.
\end{equation}
Once again, $t$ can be parametrically fixed in terms of $ \mu$ and $ \lambda$ so as to keep the ratio $ r = \Vplus / \Vtop$ constant.
For $ \mu < \mu _{c,1}$ \eqref{eq:bound1}, standard $n=1$ instantons still exist when $\ntop < 1$. Moreover,
sufficiently far from the critical point these solutions start having more negative euclidean action than HM (see Figure
\ref{fig:phaseDiagram}, left panel) and give the dominant contribution to the tunneling rate. The phase diagram on the right panel of Figure \ref{fig:phaseDiagram} describes the dominant decay channel for the class of theories \eqref{eq:potential246}. CdL instantons exist and dominate the decay rate not only when $\ntop>1$, but also for flatter barriers provided the fourth derivative of the potential at $\phitop$ is sufficiently negative to allow the presence of a new critical instanton. As we will show in Section \ref{sect:chaotic}, such critical solutions are present also for flatter potentials for which $V ^{(4)}(\phitop)=0$: therefore, the persistence and dominance of CdL solutions below $\ntop = 1$ illustrated in Figure \ref{fig:phaseDiagram} should not be regarded as a consequence of the large, negative values of  $V ^{(4)}(\phitop)$.

\begin{figure}[t]
\begin{minipage}{\smallWidthLeft}
\flushleft
\includegraphics[width=\smallWidthRight]{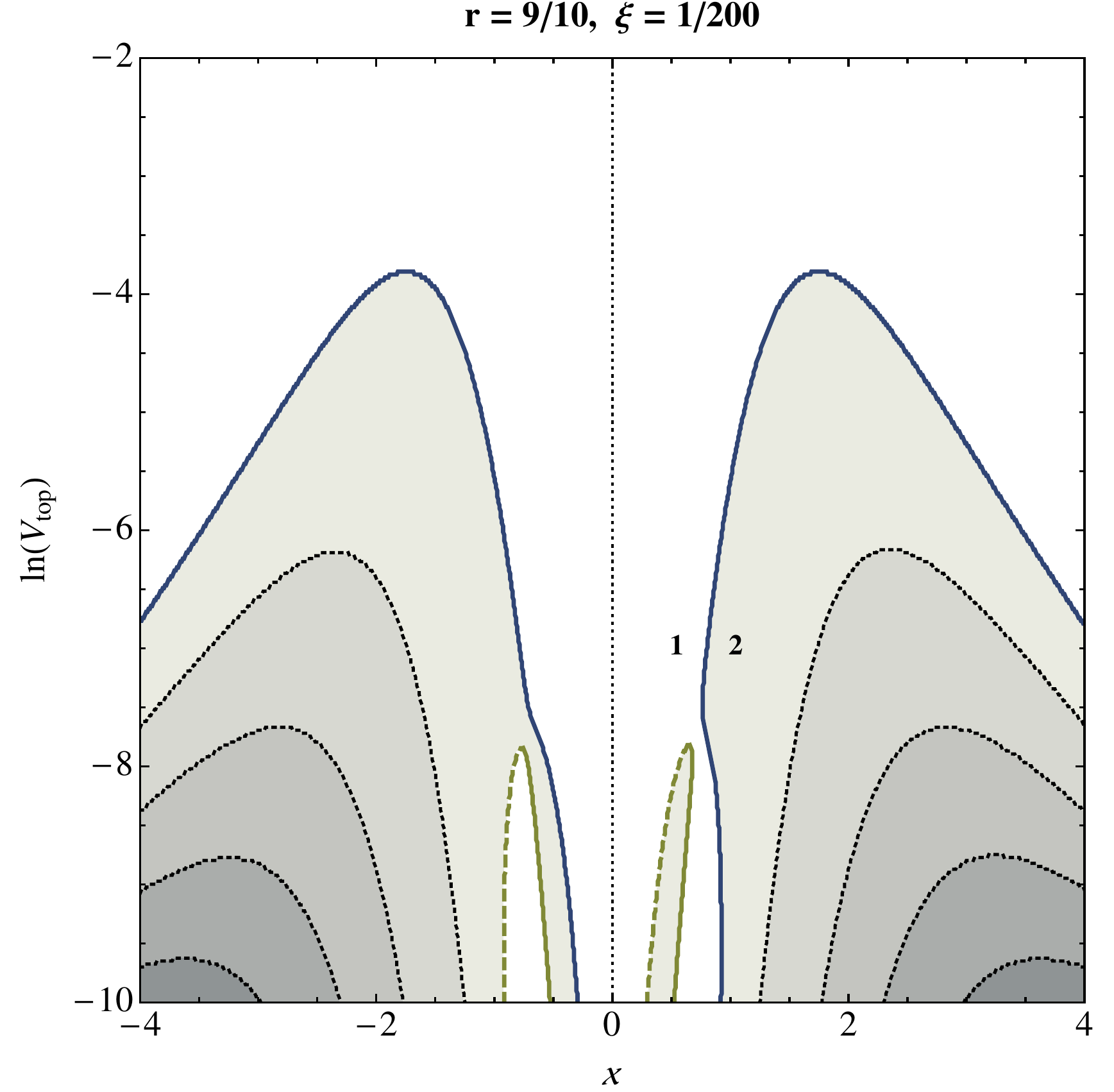}
\end{minipage}%
\begin{minipage}{\smallWidthRight}
\flushleft
\includegraphics[width=\smallWidthRight]{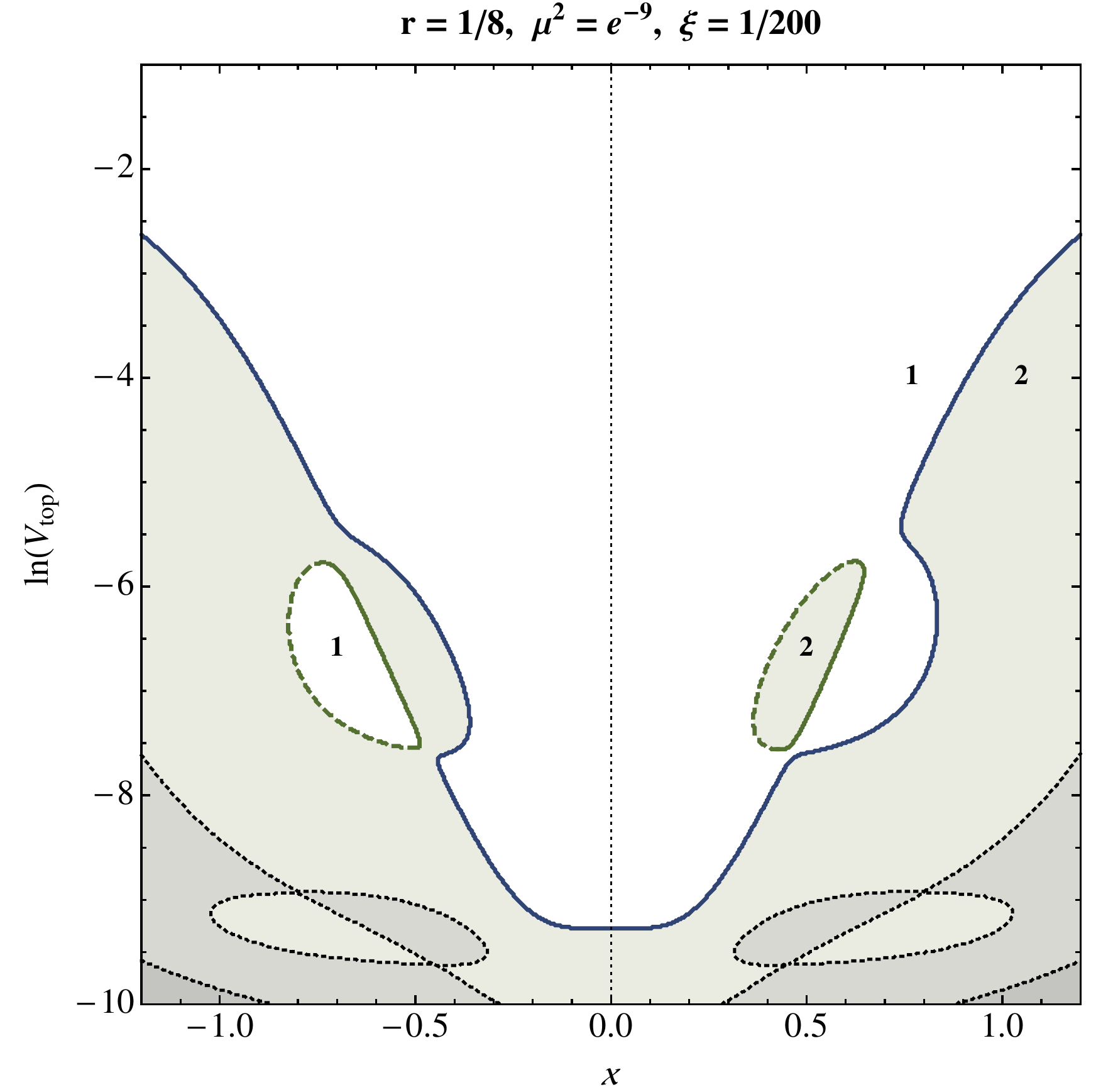}
\end{minipage}%
\caption{\label{fig:zoologyAsymmetries} Left panel: instanton diagram for the potential \eqref{eq:potential2456}, with $r=9/10$,
$ \mu ^2  = 0$ and $\xi = 1/200$. For the $n=1$ solutions, the scalar field varies between the
values at $x>0$ and the corresponding values at $x<0$: different dashings represent the correspondence
between the various branches. Right panel: detail of  the instanton diagram for $r = 1/8$, $ \mu ^2 = e^{-9}$ and $ \xi = 1/200$.
The second bifurcation point also becomes a simple critical point and an ``island'' forms in the $(x, \Vtop)$ plane.}
\end{figure}

\subsection{Asymmetric potentials}
The flat or approximately flat potentials considered so far were taken to be symmetric under
$ \varphi \rightarrow \phitop - \varphi$. In this section we briefly highlight the changes to the previous pictures when a small
asymmetry is added. We consider the potential:
\begin{equation} \label{eq:potential2456}
V = \Vtop - \frac{ \mu ^2}{2} \varphi ^2 - \frac{ \lambda}{4} \varphi ^4 - \frac{ \xi\, t ^{1/2}}{5} \varphi ^5 + \frac{t}{6} \varphi ^6 \;,
\end{equation}
and once again fix $ t$ parametrically so as to keep the ratio $r = V_{+} / \Vtop$ fixed. The left panel of
Figure \ref{fig:zoologyAsymmetries}
represents the instanton diagram for the class of theories with $ \xi = 1/200$ and $r = 9/10$.
The reduced variable $x$ was defined in Eq.~(\ref{eq:xdef}) in Section \ref{sect:standardCdL}.
Interestingly, the addition of a small asymmetry causes the disappearance of the bifurcation point, which transforms
into a simple critical point. Because of the asymmetry, none of the instanton profiles is exactly symmetric under
$ \eta \rightarrow \bar{ \eta} - \eta$. This explains how the previously symmetric branch lying above the bifurcation point
now extends smoothly into the previously asymmetric branch. As the asymmetry is increased, the critical
point moves further away from the $n=1$ branch connected to the CdL--type solutions. The number of negative modes
along each of the branches is the same as in the $ \xi = 0$ case.

When a small mass term is added the second bifurcation point presented in Figure \ref{fig:detailMassTerm} also appears as a simple critical point, and an ``island''
replaces the ring--like structure observed in the symmetric case (compare the right panel of Figure \ref{fig:zoologyAsymmetries}
with the left panel of Figure \ref{fig:detailMassTerm}). As the coefficient of the mass term is increased, the size of these islands shrinks progressively until their complete disappearance.
\begin{figure}
\centering
\includegraphics[width=\hugeWidth]{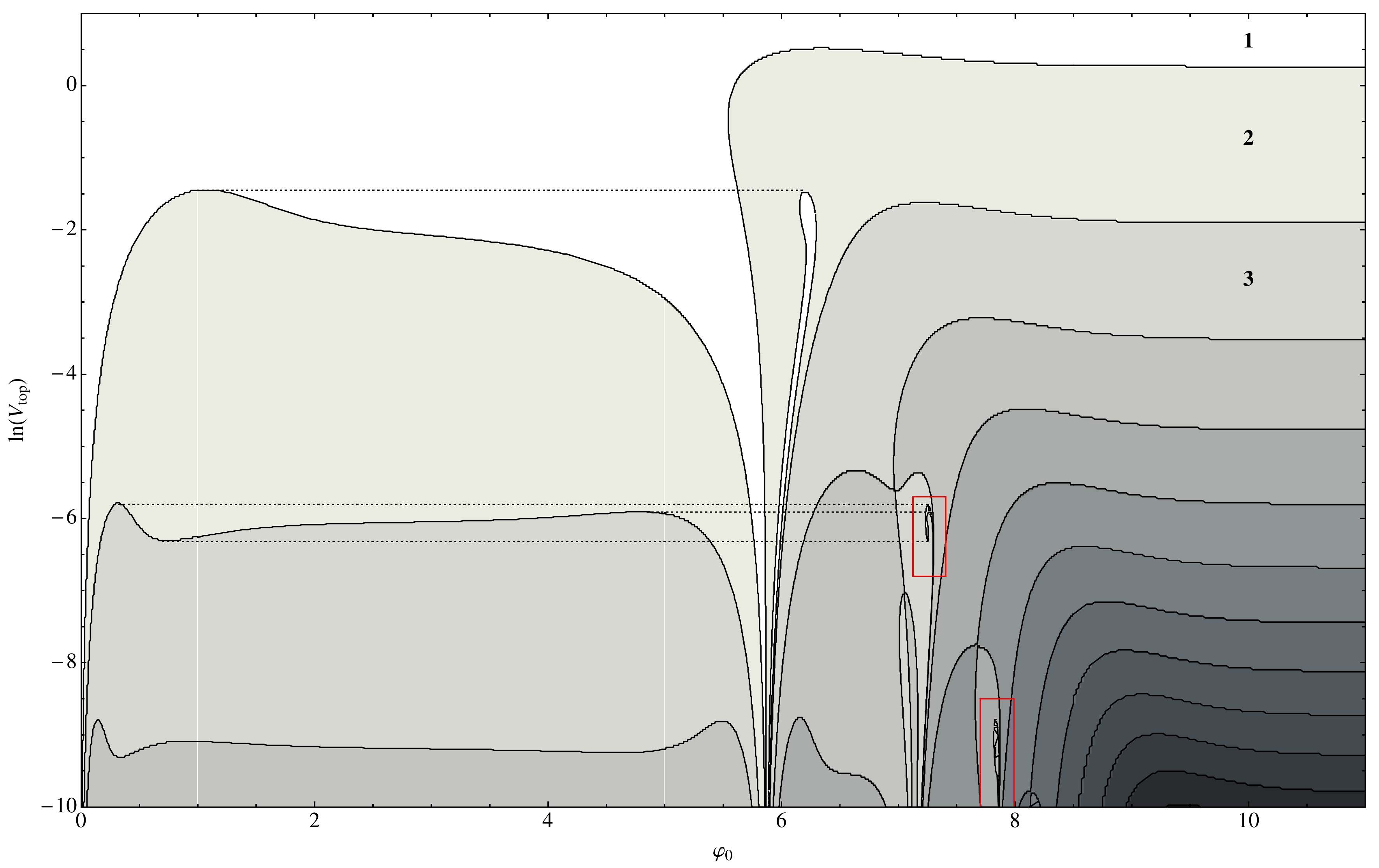} \vspace{0.5cm}\\
\centering
\includegraphics[width=\hugeWidth]{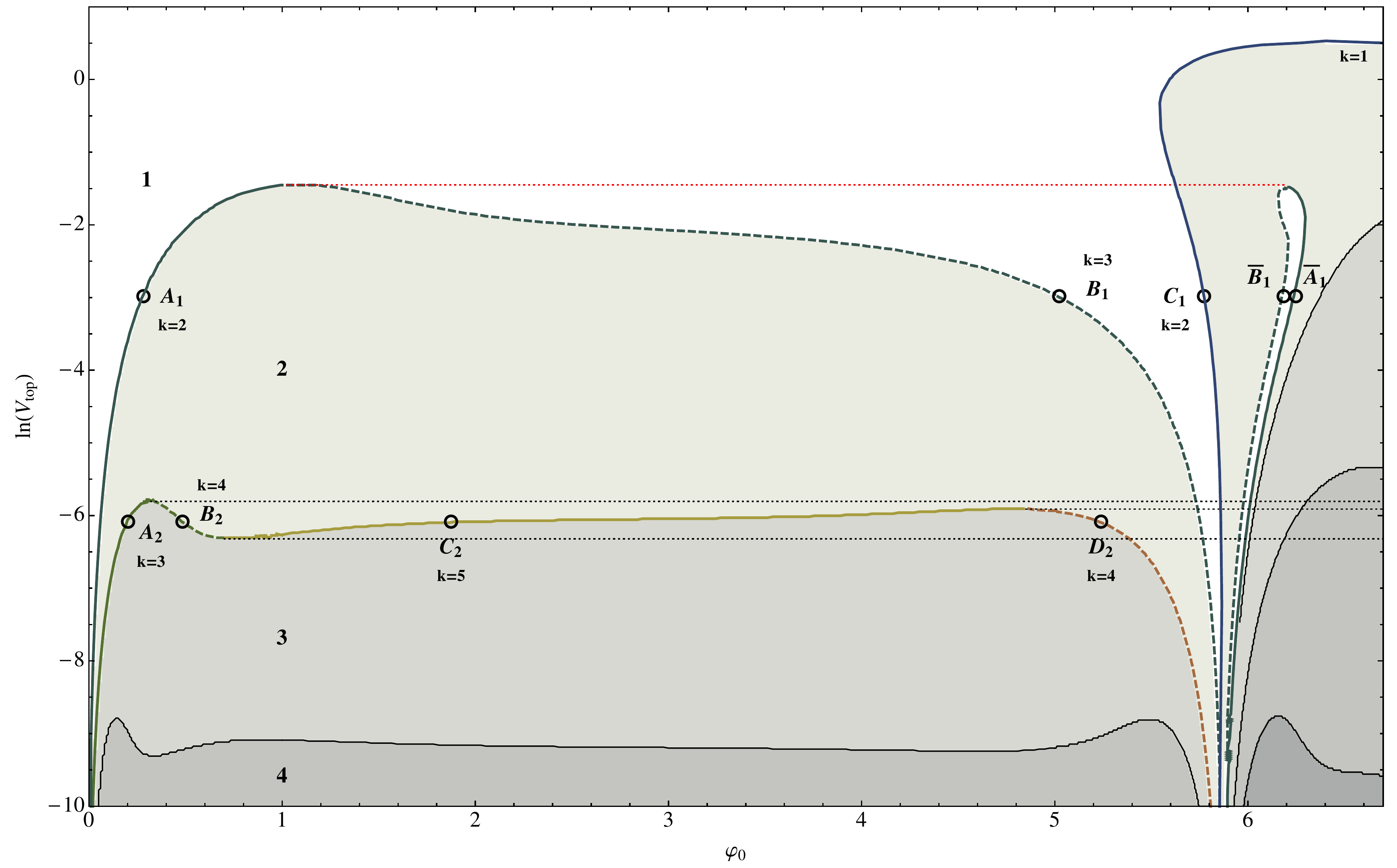}\caption{ \label{fig:phiSix}
Two views of the instanton diagram for the potential  \eqref{eq:potential6}, $ \kappa = \lambda = 1$. The two
regions higlighted in the top panel correspond to the diagrams in Figure \ref{fig:phiSixDetail}. The highlighted
points and the dashings in the bottom panel correspond to the same elements in the left panel of Figure
\ref{fig:phiSixDetail}. The number of negative modes for some branches is also indicated.
}
\end{figure}

\begin{figure}[]
\begin{minipage}{\smallWidthLeft}
\flushleft
\includegraphics[width=\smallWidthRight]{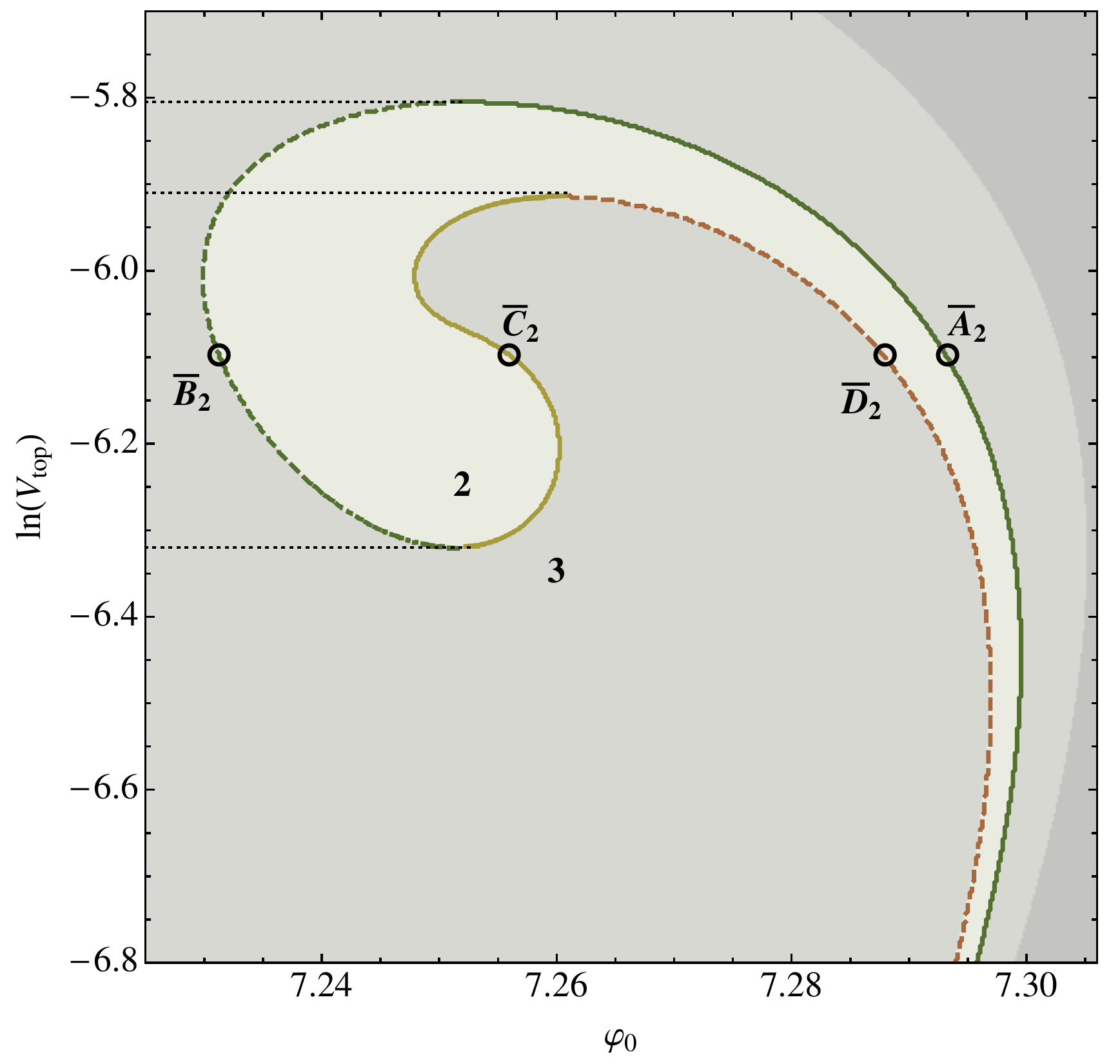}
\end{minipage}%
\begin{minipage}{\smallWidthRight}
\includegraphics[width=\smallWidthRight]{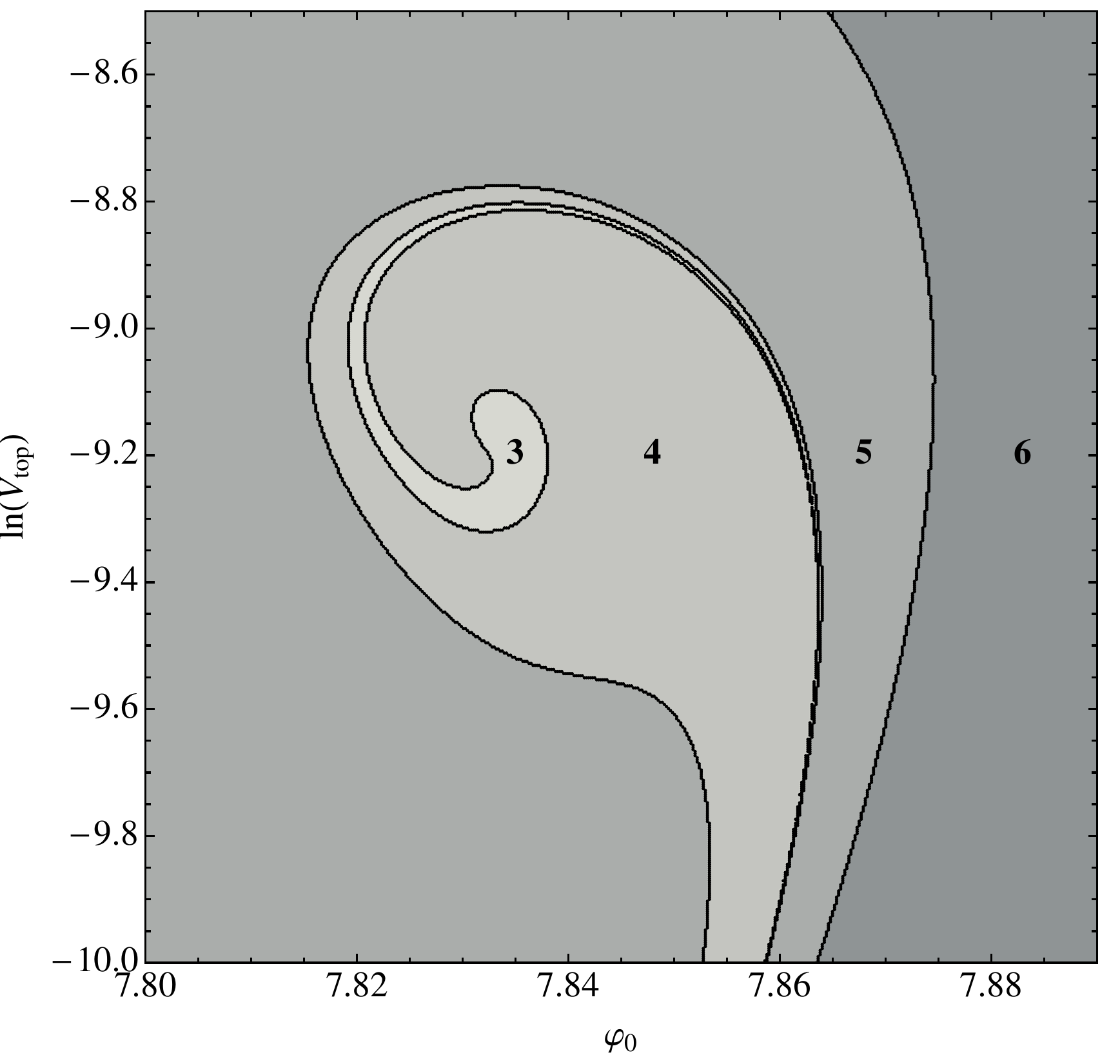}
\end{minipage}
\caption{ \label{fig:phiSixDetail} Two details of the instanton diagrams in Figure \ref{fig:phiSix}.
The highlighted points correspond to those in the bottom panel of Figure \ref{fig:phiSix} under the transformation
$ \varphi( \eta) \rightarrow - \varphi( \bar{ \eta} - \eta)$.}
\end{figure}\mbox{}

\section{Higher order flatness: $V = \Vtop - \frac{1}{6} \varphi ^6$ \label{sect:chaotic}}
As we have seen, the existence of instantons for potentials with $\ntop < 1$ can be related, in the vicinity of the
critical HM solutions, to the presence of a large and negative fourth derivative term in the expansion of the potential near
$\phitop$. To complete our overview of instanton solutions in the presence of flat potential barriers, we studied a class of
potentials for which $V ^{(4)}_{top}$ vanishes, and the potential barrier is even flatter:
\begin{equation} \label{eq:potential6}
V = \Vtop - \frac{ \lambda}{6} \varphi ^6 \;.
\end{equation}
By an appropriate choice of units one may again set $ \lambda = \kappa = 1$. Figures \ref{fig:phiSix} and \ref{fig:phiSixDetail}
present different regions of the instanton diagram corresponding to the family of theories \eqref{eq:potential6}.

From the very rich structure of the diagram we can extract the following observations:
\begin{itemize}
\item at large $\phizero$, the approximately shift-symmetric pseudo--inflationary solutions in the euclidean potential $ -V$
are again present. These solutions are exactly symmetric across $\varphi=0$, and have as many negative
modes as oscillations ($k=n$).
\item many additional critical instantons appear. Correspondingly, new branches at smaller $\phizero$
are present for every visible value of $n$. Most of these branches, just as
in the case of a $-\varphi^4$ potential, correspond to asymmetric solutions. For example, the $n=1$ solutions $A_1$ and $B_1$ correspond
to $ \bar{A}_1$ and $ \bar{B}_1$ under the transformation $ \varphi( \eta) \rightarrow - \varphi( \bar{ \eta} - \eta)$.
\item several bifurcation points are visible on the $n=3$ and $n=5$ curves in the top panel of Figure \ref{fig:phiSix}.
\end{itemize}
The most surprising feature emerging from the instanton diagrams is the chaotic character of the landscape of solutions:
for example, the solutions corresponding to the rather separated points $ A_1$ and $B_1$ end on the field values corresponding
to $ \bar{A}_1$ and $ \bar{B}_1$. Furthermore, as $ \Vtop$ decreases the distance in field space between $A_1$ and $B_1$
 remains approximately constant, while $ \bar{A}_1$ and $ \bar{B}_1$ approach each other. The
 chaotic behavior is displayed even more clearly by the $n=2$ solutions $A_2$, $B_2$, $C_2$ and $D_2$. These
 represent further asymmetric solutions, and their conjugated points lie on the very small structure depicted in the left panel
 of Figure \ref{fig:phiSixDetail}. A similar structure appears for the $n=3$ solutions (Figure \ref{fig:phiSixDetail}, right panel).

These result indicate that a rich structure of instanton solutions in theories with $\ntop < 1$ can be found without the
presence of a large fourth derivative term at $ \varphi= \phitop$. Moreover, when the potential is dominated by a $ - \varphi^6$
term around the top of the barrier, the non--linearity of the field equations allows a very complicated landscape of solutions which,
however, are not directly relevant for vacuum decay, having more than one negative mode and larger euclidean action
than HM. As a final application and in the same spirit as in Section \ref{sect:realistic}, we checked that if a regularization is added so as to make the potential positive definite,
\begin{equation}
V = \Vtop - \frac{ \lambda}{6} \varphi ^6 + \frac{ t}{8} \varphi ^{8} \;,
\end{equation}
the complex structure of the instanton diagram in Figure \ref{fig:phiSix}
does not disappear.

\section{Concluding Remarks}\label{sect:concl}

Our systematic investigation of instantons  in Einstein gravity coupled to a single self-interacting scalar field theory
with relatively flat potentials shows the existence of a very complex landscape of solutions.
We studied in detail the solutions appearing with quartic and higher order potentials, with and
without mass term, and investigated the influence of potential regularizations and asymmetries on the instantons properties.

Our study highlights the role of critical instantons as keystones for the existence of non--standard branches of
solutions. The existence of one or more critical instantons appears as a general feature of flat or almost flat potentials and, as our
results for the sextic potential illustrate, does not depend on a large negative value of the fourth derivative of the potential at the top. The new branches that emanate from these critical points typically contain more than one negative mode in their spectrum of fluctuations, indicating that one may expect solutions with fewer negative modes and lower action nearby. We have found that, in the case that the potential is regularized so as to contain a minimum at positive values of the potential, ordinary CdL instantons always persist near these minima, and they typically provide the dominant contribution to the decay rate out of these vacua.

Close to the top of the potential barrier, critical instantons are also associated to new instanton branches with a small field excursion and additional negative modes,
which connect to the critical HM solution when a small curvature term is added to the potential. The non--linear character of these new branches is indicated by the presence of bifurcation points, which we showed to be
related to the exact symmetry of the potential, and of a very complex space of solutions in the
$V = \Vtop - \frac{1}{6} \varphi ^6$ case.
We showed that these features appear regardless of the positive definiteness of the scalar potential.

There are several open questions related to our research.
An obvious one is to ask whether or not the complexity remains for potentials with an even higher degree of flatness -- we strongly suspect that it does.
Furthermore, given that our investigation was mainly numerical, it would be interesting to see if one could obtain some of these instanton solutions analytically. For instance, it may perhaps be possible to determine the critical instantons analytically as fixed points of a corresponding dynamical system.

Instanton diagrams play a central role in our results. With the more standard techniques, the rich spectrum of solutions we found would look incoherent and lacking of a precise structure. The key advantage of instanton diagrams is to single out critical instantons, whose existence determines the global structure of the space of solutions. A natural development of our work will be to attempt a more systematic analysis of extensions to theories with additional parameters, as well as extensions to multi-field \cite{Cvetic1995,Johnson2008,Yang2010,Aguirre2010,Brown2010,Ahlqvist2011} and reduced-symmetry instantons \cite{Masoumi2012}.

The most important question is probably the issue of what the correct interpretation and physical relevance of the new branches of solutions that we uncovered are. In particular, it would be interesting to know whether the new branches contribute to metastable vacuum decay.
There are two extreme viewpoints on this subject. According to the ``orthodox'' view \cite{Coleman:1987rm}, euclidean
solutions with more than one negative mode have nothing to do with tunneling. However, one should bear in mind that this statement was only proven in the absence of gravity, which leaves some room for speculation.  Recently in the context of oscillating instantons a more
"heretic" viewpoint  was discussed in \cite{Lee:2012qv}, where it was conjectured that the existence of numerous
solutions with various numbers of negative modes might sum up in the functional integral and give a contribution to decay rates.
An interesting challenge for future investigations will be to clarify these issues, and to determine the physical role of these intricate solutions.

\section*{Acknowledgements}

\par
G.L. is thankful to the Quantum Gravity group of the Albert-Einstein-Institute
and especially Hermann Nicolai for kind hospitality during his visit to Potsdam.
L.B. and J.L.L. gratefully acknowledge the support of the European Research Council
via the Starting Grant numbered 256994.

\appendix
\section{Singularities in instanton solutions \label{app:singularities}}
\subsection{No singularities at finite \texorpdfstring{$\rho$}{TEXT} \label{app:one}}
First, we prove that, if the potential $V$ is regular everywhere, solutions of (\ref{eq:scalarFieldx}, \ref{eq:conservationx})
cannot become singular at a finite value of $ \rho$. This is a slight generalization of the proof presented in \cite{Bousso2006},
where the potential was taken to be bounded. Let $n( \eta) \equiv \log{ \rho}( \eta)$ be well--defined for $ \eta < 0$ so that $ \eta = 0$
 is the candidate singular point. To facilitate reading we reproduce the field equations in both their forms:
\begin{eqnarray}
\varphi'' & = & - 3 \frac{ \rho'}{ \rho} \varphi' + V_{, \varphi} = - 3 \nt' \varphi' + V_{, \varphi} \;, \label{eq:appone}\\
\frac{ \rho ^{ \prime 2}}{ \rho ^2} & = & \frac{1}{ \rho ^2} + \frac{ \kappa }{3} \left( \frac{1}{2} \varphi ^{ \prime 2} - V \right) \;,
\label{eq:apptwo}\\
\nt ^{ \prime 2} & = &e^{ - 2 \nt} + \frac{ \kappa}{3} \left( \frac{1}{2} \varphi ^{ \prime 2} - V \right)  \;, \label{eq:appthree}\\
\nt'' & = & - e^{- 2 \nt} - \frac{ \kappa \varphi ^{ \prime 2}}{ 2}  \label{eq:appfour}\;.
\end{eqnarray}
From \eqref{eq:appfour} we know that $\nt( \eta)$ is a concave function. Therefore, only two cases
are possible:
\begin{enumerate}
\item $\nt'$ approaches a finite constant as $ \eta \rightarrow 0 ^{-}$,
\item $\nt' \rightarrow -  \infty$ as $ \eta \rightarrow 0 ^{-}$.
\end{enumerate}
In the first case, the singularity must show up in the behavior of the scalar field $ \varphi$. However, possible divergences
in $V$ and $ \varphi'$ are then related by \eqref{eq:appthree}:
\begin{equation}
\lim_{ \eta \rightarrow 0 ^{-}} \left( \frac{1}{2} \varphi ^{ \prime 2} - V \right) = \nt _0 ^{ \prime 2} - e^{- 2 \nt _0}  \;,
\end{equation}
where zero indices denote quantities evaluated at $ \eta = 0$. In the vicinity of $ \eta = 0$, the equation for the scalar field is that
of a particle in a potential $-V$ with a friction force proportional to its speed, and constant coefficient $ 3 \nt_0'$. Therefore,
a singularity in the scalar field can only be driven by the potential diverging as $ \varphi \rightarrow \pm \infty$. However,
the divergence of $ \varphi$ requires $ \varphi'$ to diverge at least as $ \eta ^{-1}$. Equation \eqref{eq:appfour} then implies
that $ \nt'$ is also divergent near $ \eta = 0$,
\begin{equation}
\nt'' \leq  - \frac{ \kappa \varphi ^{ \prime 2}}{ 2} \propto - \frac{1}{ \eta ^2} \;,
\end{equation}
which contradicts our initial hypothesis. In other words, if the scalar field diverges
due to an instability of the euclidean potential $-V$, $ \rho'/ \rho$ also diverges near the singular point. Now, we want to exclude
the possibility that this divergence could take place at finite $ \rho$.

Let's suppose now that $ \nt \rightarrow \nt_0 > 0$ while $ \nt' \rightarrow - \infty$. Because $n$ is a convex function, we can
assume the asymptotic behavior
\begin{equation}
\nt \simeq \nt_0 + \alpha\,( - \eta) ^{p}, \quad 0<p<1 \;.
\end{equation}
Inserting this behavior in \eqref{eq:appfour}, we find
\begin{equation} \label{eq:divergencePhi}
| \varphi'| \sim ( - \eta) ^{-1 + p/2} \;.
\end{equation}
This can be plugged back into \eqref{eq:appone}, resulting in
\begin{equation}
|V_{, \varphi}| \sim ( - \eta) ^{-2 + p/2} \;.
\end{equation}
For a regular potential this kind of divergence is only possible as $ \varphi \rightarrow \pm \infty$, which is however not
attainable when $ \eta \rightarrow 0 ^{-}$ because \eqref{eq:divergencePhi} implies that the scalar field ``speed'' $ \varphi'$ does not diverge fast enough.
\subsection{Singular instantons and runaway of the scalar field \label{app:two}}
Following the results of Appendix \ref{app:one}, only compact solutions of the equations for $O(4)$--invariant instantons can
be singular. Moreover, the singularities are all characterized by
\begin{equation}
\rho \stackrel{ \eta \rightarrow 0 ^{-}}{ \longrightarrow} 0 \;,
\end{equation}
taking $ \rho$ to be regular for $ \eta < 0$. We now show the relation between the behavior of the scalar field and of the metric function $ \rho$ near the point $ \rho = 0$,
both in the singular and non--singular case.

Suppose first that $ \varphi \rightarrow \phizero$ as $ \eta \rightarrow 0 ^{-}$, where $ \phizero$ is some finite value.
In this case, the derivative combination $ \frac{ \rho'}{ \rho} \varphi'$ must remain finite, otherwise the asymptotic
scalar field equation \eqref{eq:appone} would read
\begin{equation}
\varphi'' \simeq - 3 \frac{ \rho'}{ \rho}  \varphi' \quad \Rightarrow \quad \varphi' \propto \rho ^{-3} \;,
\end{equation}
In this case, the asymptotic form of \eqref{eq:apptwo} reads
\begin{equation}
\rho ^{ \prime 2} \propto \rho ^{ -4} \quad \Rightarrow \quad \rho \propto ( - \eta) ^{1/3} \;.
\end{equation}
In this case, however, $ \varphi' \propto ( - \eta) ^{-1}$ and $ \varphi$ diverges logarithmically, which contradicts our initial
hypothesis. Therefore, if the solution is such that $ \varphi$ approaches a finite value, the combination $ \frac{ \rho'}{ \rho} \varphi'$
must approach a finite value. As $ \rho' / \rho$ is diverging monotonically, this requires $ \varphi'$ to approach zero, in which
case one also finds from \eqref{eq:apptwo}
\begin{equation}
\rho ' \stackrel{ \eta \rightarrow 0 ^{-}}{ \longrightarrow} 1 \;.
\end{equation}
This means that whenever $ \varphi$ approaches a finite constant the solution is actually fully non--singular.

The nature of the singularity appearing in compact solutions with $ \varphi$ diverging as $ \rho \rightarrow 0$ might depend
on the asymptotics of the scalar field potential. However, for potentials which are bounded or which diverge (positive or negative)
 with a power law, one can easily check that
\begin{eqnarray}
\rho & \propto & ( \bar{ \eta}  - \eta) ^{1/3} \;, \\
\varphi & \propto & \log{ \left(\bar{ \eta} - \eta \right)} \;,
\end{eqnarray}
is a an asymptotic solution of the field equations. From the point of view of the scalar field equation \eqref{eq:appone},
this means that provided the potential is not too steep the divergence of $ \varphi$ is always asymptotically driven by the
anti--friction term $ - 3 \frac{ \rho'}{ \rho} \varphi'$.

\section{Continuous families of solutions and perturbation modes\label{app:three}}
In Section \ref{ssect:diagrams} we stated that when the instanton curve $\Vtop( \phizero)$
has a stationary
point at $ \phizero ^{*}$, the solution corresponding to this value of $ \phizero$ possesses a regular perturbation mode.
In this section we prove this statement in the simplified context of a theory with no reparametrization invariance.

Let $S[ \varphi(x), p]$ be an action which depends \textit{explicitly} on some real parameter $p$. Let now $ \varphi_ \lambda(x)$ be
a family of solutions for the family of theories specified by $S[ \varphi, p( \lambda)]$. Finally, let's assume $ \varphi_ \lambda(x)$
to be a regular function of $ \lambda$. We prove that, if $ p'( \lambda ^{*}) = 0$, then
$$
\partial _{\lambda} \varphi(x) \Big|_{ \lambda = \lambda ^{*}}
$$
is a regular perturbation mode of $ \varphi_{  \lambda ^{*}}$. Let $ \delta \varphi$ be a generic field variation.
We now assume $ t$ to be infinitesimal and consider the quantity
\begin{eqnarray*}
S\Big[ \varphi_{ \lambda ^{*} + \delta \lambda} + t \, \delta \varphi, p(\lambda ^{*} + \delta \lambda)\Big] & = &
S\Big[ \varphi _{ \lambda ^{*}} + t \, \delta \varphi + \delta \lambda ( \partial _{\lambda} \varphi _{\lambda})|_{ \lambda =\lambda ^{*}},
p( \lambda ^{*}) \Big] + \mathcal{O}( \delta \lambda ^2) \\
& = & S[ \varphi  _{ \bar{\lambda}}, p( \lambda ^{*})] + \mathcal{O}( t ^2, \delta \lambda ^2) \\
&& + t \, \delta \lambda\, \int d ^4x\, d ^4y\, \delta \varphi(x) \frac{ \delta ^2 S [ \varphi _{ \lambda ^{*}}, p( \lambda ^{*})] }{
\delta \varphi(x) \delta \varphi(y)}( \partial _{\lambda} \varphi _{\lambda})|_{ \lambda =\lambda ^{*}}(y) \;.
\end{eqnarray*}
Of course, no terms in $t$ and $ \delta \lambda$ are present because of the field equations for $ \varphi_{\lambda ^{*}}$.
On the other hand, the field equation for $ \varphi_{ \bar{\lambda} + \delta \lambda}$ implies
\begin{equation}
\frac{dS[ \varphi_{ \lambda ^{*} + \delta \lambda} + t \, \delta \varphi, p( \bar{\lambda} + \delta \lambda)\Big]}{dt} \Big|_{t=0} = 0 \;.
\end{equation}
At order $ \delta \lambda$, this implies
\begin{equation}
\int d ^4x\, d ^4 y\, \delta \varphi(x) \frac{ \delta ^2 S[ \varphi _{ \lambda ^{*}}, p( \lambda ^{*})] }{ \delta \varphi(x) \delta \varphi(y)}
( \partial _{ \lambda} \varphi_ \lambda)|_{ \lambda =\lambda ^{*}}(y) = 0 \;.
\end{equation}
As $ \delta \varphi(x)$ is a generic field variation, this implies that $\left.( \partial _{\lambda} \varphi _{\lambda})
\right|_{ \lambda = \lambda ^{*}}$
is a regular perturbation of $ \varphi_{ \lambda ^{*}}$
\begin{equation}
\int d ^4y \frac{ \delta ^2 S[ \varphi _{ \lambda ^{*}}, p( \lambda ^{*})] }{ \delta \varphi(x) \delta \varphi(y)} ( \partial _{ \lambda}
\varphi_  \lambda)|_{ \lambda =\lambda ^{*}}(y) = 0 \;.
\end{equation}

  \bibliographystyle{utphys}
    \bibliography{flat_potentials}
\end{document}